\documentclass[pra,aps,10pt,twocolumn,showpacs,showkeys]{revtex4-1}
\usepackage{epsfig}
\usepackage{enumitem}
\usepackage{scalefnt}
\usepackage{amsmath,amssymb,bm,bbm}
\usepackage[mathscr]{euscript}
\usepackage{upgreek}

\newcommand\paritStyle[1]{\textrm{\mdseries\upshape(\textit{#1}\kern0.1ex)}}

\newcommand\mapstoname[2][1.5em]{\DOTSB\mapstochar\xrightarrow{\parbox{#1}{\centering $\scriptstyle #2$}}}
\renewcommand\ge\geqslant
\renewcommand\le\leqslant
\newcommand\idop{\mathbbm{1}}

\makeatletter
	\newcommand\ket[1]{
			\left\lvert#1%
			\@ifnextchar\bra{\right\rangle\mspace{-5mu}}{\right\rangle}%
	}		
\makeatother
\newcommand\bra[1]		{\left\langle#1\right\rvert}		
\newcommand\gen[1]		{\left\langle #1 \right\rangle}
\newcommand\sympl[1]	{\left[ #1 \right]}
\newcommand\ens[1]		{\left\{ #1 \right\}}

\newcommand\bpar[1]{\textbf{\upshape(#1)}}

\newcommand\Z{\mathbb Z}
\newcommand\C{\mathbb C}

\newcommand\cH{\mathcal H}
\newcommand\cP{\mathcal P}

\newcommand\sS{\mathscr S}
\newcommand\sT{\mathscr T}

\newcommand\sfW{\mathsf W}
\newcommand\sfX{\mathsf X}
\newcommand\sfZ{\mathsf Z}

\newcommand\Cliff{\mathcal C}
\newcommand\sCliff{\upsigma\mathcal C}

\def\Mod[#1]{\ensuremath{\mathsf{Mod}_{#1}}}
\def\coMod[#1]{\ensuremath{\mathsf{coMod}_{#1}}}
\def\parity{\ensuremath{\oplus}}
\def\Log{\ensuremath{\mathsf L}}
\def\P{\ensuremath{\mathsf P}}

\def\BPP{\ensuremath{\mathsf{BPP}}}
\def\BQP{\ensuremath{\mathsf{BQP}}}
\newcommand\NC[1][]{\ensuremath{\mathsf{NC^{#1}}}}

\newcommand\e{\mathrm e}
\newcommand\x{\times}
\newcommand\ox{\otimes}

\newcommand\packmathbox[2]{\mbox{$#1#2$}}
\newcommand\mathbox{\mathpalette\packmathbox}

\newcommand\ctrlP{\mathbox{\Uplambda\!\!\;\textup{P}}}
\newcommand\cW{\mathbox{\Uplambda\!\!\;\textup{W}}}
\newcommand\cZ{\mathbox{\Uplambda\!\!\;\textup{Z}}}
\newcommand\cX[1][]{\mathbox{\Uplambda_{#1}\!\!\:\textup{X}}}
\newcommand\cnot{\textsc{cnot}}
\newcommand\cOp[2][]{\mathbox{\Uplambda_{#1}\!\!\:\textup{#2}}}

\def\DefStabMeas{\mbox{\scalefont{0.9}\textsc{Definite Stabilizer Measurement}}}
\def\StabMeas{\mbox{\scalefont{0.9}\textsc{Stabilizer Measurement}}}
\def\GottKnill{\mbox{\textsc{\scalefont{0.9}Gottesman-Knill}}}

\newcommand\ie{\textit{i.e.}}
\newcommand\eg{\textit{e.g.}}
\newcommand\etc{\textit{etc.}}
\newcommand\etal{\textit{et al.}}

\newcommand\decision{decision}

\DeclareMathOperator\Sp{Sp}
\DeclareMathOperator\U{U}
\DeclareMathOperator\Tr{Tr}
\DeclareMathOperator\img{img}

\newcommand\herm{^\dagger}
\newcommand\trans{^\top}

\newcommand\units{^\ast}
\newcommand\sox[1]{^{\otimes #1}}

\renewcommand\vec[1]{\bm{\mathbf{#1}}}
\newcommand\unit{\vec{\hat{e}}}

\newcounter{theorem}
\newcounter{definition}
\newcounter{newcorollary}[theorem]
\renewcommand\thedefinition{\Roman{definition}}
\renewcommand\thenewcorollary{\thetheorem\alph{newcorollary}}

\newcommand\definition{%
	\refstepcounter{definition}%
	\par
	\vspace{8pt}%
	\noindent
	{\bf Definition~\thedefinition:}}
\def\enddefinition{\par\vspace{8pt}}

\newcommand\lemma{%
	\refstepcounter{theorem}%
	\par
	\vspace{8pt}%
	\noindent
	{\bf Lemma~\thetheorem:}}
\def\endlemma{\par\vspace{8pt}}

\newcommand\theorem{%
	\refstepcounter{theorem}%
	\par
	\vspace{8pt}%
	\noindent
	{\bf Theorem~\thetheorem:}}
\def\endtheorem{\par\vspace{8pt}}

\newcommand\corollary{%
	\refstepcounter{newcorollary}%
	\par
	\vspace{8pt}%
	\noindent
	{\bf Corollary~\thenewcorollary:}}
\def\endcorollary{\par\vspace{8pt}}

\newcommand\proof{%
	\par
	\vspace{8pt}%
	\noindent
	{\bf Proof:}}
\def\endproof{\par\vspace{8pt}}
\def\QED{\hfill$\square$}

\newcommand\convention{%
	\par%
	\vspace{8pt}
	\noindent
	{\bf Convention:}}
\def\endconvention{\par\vspace{8pt}}

\newcommand\PauliMeasurePreamble{%
		For qudits of dimension $d \ge 2$, most Pauli operators are non-Hermitian.
		For a Pauli operator $P$ of order at most $d$, we 
		use the phrase ``measurement of $P$\,'' as short-hand, to refer to measurement of any Hermitian operator $H$ with a spectral diameter less than $2\pi$, such that $P = \exp(iH)$.
		This amounts to collapsing the state of the system onto one of the eigenspaces of $P$ via projective measurement, obtaining a record $h \in \Z_d$ of the eigenvalue $\tau^{2h}$ associated to that eigenspace.}

\begin{document}

	\title{A linearized stabilizer formalism for systems of finite dimension}
	\author{Niel de Beaudrap}
	\email{Contact: niel.debeaudap@gmail.com}
	\affiliation{Department of Applied Mathematics and Theoretical Physics, University of Cambridge, Wilberforce Road, Cambridge, UK}
	\keywords{stabilizer formalism, Weyl operators, Clifford group, symplectic group}
	\date{10 September 2012}

\begin{abstract}
	The stabilizer formalism is a scheme, generalizing well-known techniques developed by Gottesman~\cite{GottPhD} in the case of qubits, to efficiently simulate a class of transformations (\emph{stabilizer circuits}, which include the quantum Fourier transform and highly entangling operations) on standard basis states of $d$-dimensional qudits.
	To determine the state of a simulated system, existing treatments involve the computation of cumulative phase factors which involve quadratic dependencies.
	We present a simple formalism in which Pauli operators are represented using displacement operators in discrete phase space, expressing the evolution of the state via linear transformations modulo $D \le 2d$.
	We thus obtain a simple proof that simulating stabilizer circuits on $n$ qudits, involving any constant number of measurement rounds, is complete for the complexity class \coMod[d]\Log\ and may be simulated by $O(\log(n)^2)$-depth boolean circuits for any constant $d \ge 2$.  
\end{abstract}

\maketitle

\section{Introduction}        
Efficiently simulating the evolution of quantum states is a problem of central importance to modern physics.
It is generally considered to be a difficult problem: this is formalized by the conjecture that the complexity class $\P$, representing \decision\ problems efficiently and deterministically solvable by conventional computers, does not contain $\BQP$, the class of \decision\ problems solvable efficiently and with bounded error by using
local many-body interactions acting on a standard basis state.
 
However, some transformations of quantum states are indeed easy to classically simulate.
The \emph{stabilizer formalism} is one technique to do so, which may be applied for some evolutions of finite-dimensional systems.
Developed by Gottesman~\cite{GottPhD} to study quantum error correction on qubits, the stabilizer formalism 
describes how to simulate \emph{stabilizer circuits}: CP maps obtained by composing Clifford group operations (unitary maps which preserve the group generated by Pauli spin operators $\idop$, $\sigma_x$, $\sigma_y$, $\sigma_z$), Pauli observable measurements, and Clifford operations controlled directly on measurement outcomes~\cite{HeisQC}.
This serves as a foundation for other efficient simulation techniques~\cite{JM08,JKMW10,Nest10} and for measurement-based quantum computing~\cite{RBB03,CLN05}.
It has since been extended in similar ways by Gottesman~\cite{G98} and by Hostens, Dehaene, and de Moor~\cite{HDM05} to systems of qudits of dimension $d \ge 2$.
This generalization of the stabilizer formalism can be used to simulate circuits involving the quantum Fourier transform (an important element of Shor's integer factoring algorithm~\cite{Shor94}) and certain arithmetic operations on standard basis states (which are closely related to other key operations involved in Shor's algorithm).

The stabilizer formalism is notable in part for the minimal computational power it requires.
Aaronson and Gottesman~\cite{AG04} show that for qubits, simulating stabilizer circuits on computational 
basis states is feasible for the complexity class \parity\Log, the subclass of $\P$ consisting of those problems which are reducible (using logarithmic workspace) to solving systems of linear equations modulo~2~\cite{D90}. 
However, in existing algorithms~\cite{AG04,DM03}, the exact state of the system is determined by scalar factors which depend quadratically on the exponents of Pauli operators $\sigma_x^a \sigma_z^b$\,; similar remarks apply to the formalism of Ref.~\cite{HDM05} for $d \ge 2$.
One may ask whether these quadratic dependencies are necessary for the efficient simulation of stabilizer circuits, or if they may be removed to obtain a formalism involving only linear algebra.

This article presents a simple modification to the formalisms of Refs.~\cite{AG04,HDM05}, to simulate stabilizer circuits by solving systems of linear equations modulo $D \in \{d,2d\}$.
Using discrete Weyl operators as a matrix basis, rather than the related operators $\sigma_x^a \sigma_z^b$, we eliminate quadratic dependencies from simulations to enable Clifford group operations to be simulated entirely linearly.
Furthermore, we demonstrate a subgroup of Clifford group which forms a group representation (acting on density operators) of the symplectic group modulo $D$, simplifying the connection to symplectic operators described in Ref.~\cite{HDM05}.
This allows us to obtain direct proofs, for any constant $d \ge 2$, that simulating a constant number of measurements in a stabilizer circuit is complete for the complexity class $\coMod[d]\Log \subseteq \P$ defined by Buntrock \etal~\cite{BDHM92} generalizing \parity\Log\ to arbitrary modulus $d \ge 2$.
Consequently, all such circuits can be simulated by $O(\log(n)^2)$-depth boolean circuits.

Beyond the complexity theoretic results of this article, the formalism that we present here is practically useful for simulations of stabilizer circuits.
This (as well as further technical results on Pauli stabilizer groups in the Appendices) should be 
useful to the study of error correction~\cite{GLG10,Gheorg11} and measurement-based computation~\cite{ZZXS03} on higher-dimensional systems.

\section{Preliminaries}
\label{sec:preliminaries}

Throughout, $d \ge 2$ is a constant.
Let $\cH_d = \C^d$ with standard basis vectors $\ket{q}$ for $q \in \{0,1,\ldots,d-1\}$.
When little confusion may result, we identify $q$ with an element of $\Z_d$ (the ring of integers mod $d$).

Our formalism differs slightly from traditional descriptions of the stabilizer formalism, to achieve as uniform a description as possible while remaining consistent with conventions \eg~for the special case $d = 2$ for qubits.
We first address a non-uniformity noted by Refs.~\cite{HDM05,G98} between the cases of $d$ even and $d$ odd.
In analogy to the case $d = 2$, we define unitary operators $X$ and $Z$ by
\begin{align}
	\label{eqn:defXandZ}
		X \ket{q} =&\; \ket{q+1}
	&
		\text{and}
	&&
		Z \ket{q} =&\; \e^{2\pi iq/d} \ket{q}	\,,
\end{align}
using addition mod $d$; then $X$ and $Z$ have order $d$.
We also wish to have an operator $Y$ which has order $d$ such that $XYZ = \tau\idop$, where $\tau^2 = \e^{2\pi i/d}$ (generalizing the equation $XYZ = i\idop$ which holds for qubits).
One may show that $(X\herm Z\herm)^d = \idop$ for $d$ odd, and $(X\herm Z\herm)^d = -\idop$ for $d$ even: the order of $\tau$ must then depend non-trivially on the parity of $d$.
We choose $\tau = \e^{i \pi (d^2+1)/d} = (-1)^d\e^{i\pi/d}$, which is consistent with $\tau = i$ for qubits~\footnote{
	This differs from $\tau = -\exp(i\pi/d)$ in Appleby~\cite{Appleby05}: these have similar features, but differ by a sign for $d$ even.
	We choose $\tau$ so that $Y = i^\dagger\! Z^\dagger\! X^\dagger$ is a Weyl operator for $d = 2$.
};
then 
$(\tau X\herm Z\herm)^d = \idop$ for all $d$.

\definition\
		 $\tau = \e^{i \pi (d^2+1)/d}$, let $X$ and $Z$ be defined as in Eq.~\eqref{eqn:defXandZ}, and $Y = \tau X\herm Z\herm$.
	The \emph{Pauli group} over $\Z_d$ is the group $\cP_d = \gen{X,Y,Z} = \gen{\tau \idop, X, Z}$.
	The $n$-qudit Pauli group $\cP_d\sox{n}$ is the group generated by 
	$n$-fold tensor products of $\idop$, $X$, $Y$, and $Z$.
\vspace*{8pt}

\noindent 
To reduce computations for phases in the stabilizer formalism, 
we describe Pauli operators in terms of translation operators in discrete phase space~\cite{Appleby05,GE08}, rather than directly as multiples of $Z^a X^b$:

\definition\
	\label{def:WeylOptor}
	A \emph{Weyl operator} (on a single qudit) is an operator $
	 		W_{a,b}
			\,=\,
			\tau^{-ab} Z^a X^b 
	$ for some $a,b \in \Z$\,.
	On $n$ qudits, for vectors $\vec a, \vec b \in \Z^n$, the Weyl operators are
	\begin{align}
			W_{\vec a, \vec b}
		&=
			W_{a_1,b_1} \ox \cdots \ox W_{a_n, b_n}
		\notag\\&=
			\tau^{-\vec a \cdot \vec b} \bigl(Z^{a_1} \ox \cdots \ox Z^{a_n}\bigr) \bigl(X^{b_1} \ox \cdots \ox X^{b_n}\bigr),
	\end{align}
	where $\vec a \cdot \vec b$ is the dot product. 
	For block vectors $\vec v = \vec a \oplus \vec b \in \Z^{2n}$, we write $W_{\vec v} = W_{\vec a, \vec b}$.
\vspace*{8pt}

\noindent 
The Pauli operators $\ens{\idop,X,Y,Z}$ are all Weyl operators (for example, $Y = W_{-1,-1}$); and the set of Weyl operators is invariant under cyclic permutations of $(X,Y,Z)$ in Definition~\ref{def:WeylOptor}.
Weyl operators have convenient properties: define the \emph{symplectic inner product} for $\vec v, \vec w \in \Z^{2n}$ by
\begin{align}
	\label{eqn:symplecticInnerProd}
		\sympl{\vec v, \vec w}
	&=
		\vec v\trans \!\sigma_{2n} \vec w,
	&
	\text{where} ~~
		\sigma_{2n}
	&=
		\left[\,\begin{matrix}
			\,\:\;0_n				&		\idop_n
		\\
			\!-\idop_n	&		0_n
		\end{matrix}\,\right]	\,.
\end{align}
Note that $\sympl{\vec v, \vec w} = -\sympl{\vec w, \vec v}$, and $\sympl{\vec v, \vec v} = 0$. 
The following may be verified by considering actions on the standard basis, using the relation $X^b Z^a = \tau^{-2ab} Z^a X^b$ for $a,b \in \Z$:
\lemma\
	\label{lemma:WeylCalculus}%
	For any $\vec v, \vec w \in \Z^{2n}$, we have $W_{\vec v} W_{\vec w} \,=\, \tau^{\sympl{\vec v, \vec w}} W_{\vec v + \vec w}$.
	The following properties then also hold:
	\begin{tabular}{c@{~~}l}
	\\[-2ex]
	\paritStyle{i}		& $W_{\vec v} W_{\vec w} = \tau^{2\sympl{\vec v, \vec w}} W_{\vec w} W_{\vec v}.$		\\[0.5ex]
	\paritStyle{ii}		&	$[W_{\vec v}, W_{\vec w}] = 0$ if and only if $[\vec v, \vec w] \equiv 0 \pmod{d}$.	\\[0.5ex]
	\paritStyle{iii}	&	$W_{\vec v}^t = W_{t \vec v}$ for $t \in \Z$; in particular, $W_{\vec v}\herm = W_{-\vec v}$. \\[0.5ex]
	\paritStyle{iv}		&	The order of $W_{\vec v}$ divides $d$.	\\[-0.5ex]
	\end{tabular}
\endlemma
\noindent
These properties are a straightforward generalization of observations made by Appleby~\cite{Appleby05} in his nearly identical formalism for single qudit operators; they are also essentially what allow us to define the formalism of this article.
They also suggest a conven\-tion of evaluating each vector modulo $d$: however, this leads to inconsistencies for $d$ even.
For instance, we have
\begin{align}
	\label{eqn:mod-d-negation}
		W_{0,1}
	\;=\;
		Z^0 X^1
	\;=\;
		\tau^{2d} Z^d X^1
	\;=\;
		-W_{d,1} \,;
\end{align}
as $W_{d,1}$ is non-zero, we then cannot equate $W_{0,1}$ to $W_{d,1}$\,.
Despite this, it is still possible to use modular arithmetic with the Weyl operators, as outlined in the next lemma:
\lemma\
	\label{lemma:WeylModular}%
	For all $\vec v, \vec w \in \Z^{2n}$, we have $W_{\vec v} \propto W_{\vec w}$ if and only if $\vec w = \vec v + d \vec x$ for some $\vec x \in \Z^n$, in which case
	\begin{align}
		\label{eqn:WeylModular}
		W_{\vec w} \,=\, (-1)^{(d+1)\sympl{\vec v, \vec x}} \:\! W_{\vec v}  .
	\end{align}
	In particular, $W_{\vec v} = W_{\vec w}$ if
		$\vec v \equiv \vec w \pmod{d}$ for $d$ odd
		, and if
		$\vec v \equiv \vec w \pmod{2d}$ for $d$ even.
\endlemma
\noindent
This follows from the action of $W_{\vec v} W_{\vec w}\herm$ on the standard basis.
We may subsume the cases of $d$ even and $d$ odd into a single formalism by adopting the following convention:

\convention\
	For $d \ge 2$, we define
	\begin{gather}
	 		D
		\;=\;
			\text{the order of $\tau$}
		\;=\;
			\begin{cases}
			 	d	&	\text{if $d$ is odd},	\\
				2d	&	\text{if $d$ is even};
			\end{cases}
	\end{gather}
	then the mapping $\vec v \mapsto W_{\vec v}$ is well-defined for $\vec v \in \Z_D^{2n}$\,, and all associated arithmetic may be performed mod $D$.
\endconvention
\noindent
We frequently describe operations in terms of arithmetic modulo $D$ --- with a notable exception:
as standard basis states $\ket{q} \in \cH_d$ involve integers $q \in \{0,1,\ldots,d-1\}$ by definition, we require expressions involving $q$ occurring as powers of $\tau$ to be invariant under replacing $q$ with $q \pm d$.

Consider the Hilbert-Schmidt inner product $
 		\langle A, B \rangle
	=
			\Tr(A\herm B)
/
			d^n
$,
renormalized
to obtain $\langle \idop, \idop \rangle = 1$.
From the the effect of $P \in \cP_d\sox{n}$ on standard basis states, one may easily show $\Tr(P) \ne 0$ if and only if $P \propto \idop$.
Then for $P, Q \in \cP_d\sox{n}$, either $P \propto Q$ or $\langle P, Q \rangle = 0$, so that:
\lemma\
	\label{lemma:weylOrthogonal}
	The following are equivalent for $\vec v, \vec w \in \Z^{2n}$, for any system of $n$ qudits of dimension $d$:
	\vspace*{-3pt}
	\begin{align*}
	  \begin{tabular}{c@{~~}l}
		\paritStyle{i}		&		$\langle W_{\vec v} , W_{\vec w} \rangle \ne 0$;\vspace{3pt}		\\
		\paritStyle{ii}		&		$\langle W_{\vec v} , W_{\vec w} \rangle = \pm 1$;\vspace{3pt}	\\
		\paritStyle{iii}	&		$W_{\vec v} \propto W_{\vec w}$;\vspace{3pt}
		\end{tabular}&\qquad&\begin{tabular}{c@{~~}l}
		\paritStyle{iv}		&	$W_{\vec v} = \pm W_{\vec w}$;\vspace{3pt}	\\
		\paritStyle{v}		&	$\vec v \equiv \vec w \!\pmod{d}$.
	  \end{tabular}
	\\[-5ex]
	\end{align*}%
	\noindent
	Furthermore, when these hold for $d$ odd, $W_{\vec v} = W_{\vec w}$\,.
\endlemma
\pagebreak
From this, it is easy to show that a single Weyl operator $W_{\vec p}$ can only be generated by a collection of other Weyl operators $W_{\vec v_1}, \ldots, W_{\vec v_\ell}$ if $W_{\vec p} = \pm W_{\vec v_j}$ for some operator $W_{\vec v_j}$.
By a simple dimension-counting argument, and using the fact that two integers $0 \le a,b < d$ are equivalent mod $d$ if and only if they are equal, we may show:

\corollary\
 	The operators $W_{\vec v}$ for $\vec v \in \{0,\ldots,d-1\}^{2n}$ form an orthonormal basis for the linear operators on $\cH_d\sox{n}$ with the inner product 
	$\langle A, B \rangle = \Tr(A\herm B)/d^n$.
\endcorollary

\noindent
Note that the Weyl operators $W_{\vec v}$ for $\vec v \in \Z_D^{2n}$ are \emph{not} independent for $D$ even; this is an important technical point in the analysis to follow.


\section{Simulating unitary stabilizer circuits}
\label{sec:simulateClifford}

We now describe a simple formalism, based on Weyl operators, for simulating an important subclass of stabilizer circuits: the class of unitary Clifford circuits, acting on standard basis states, and including at most one final measurement.
We refer to these as \emph{unitary stabilizer circuits}.

Note that unitary stabilizer circuits exclude intermediate measurements, or Clifford operations controlled by measurement outcomes.
If we are only concerned with the distribution of a single measurement as output, we can still simulate circuits with classically controlled \emph{Pauli} operations conditioned on measurement outcomes, by applying the principle of deferred measurement (as we outline in Section~\ref{sec:princDeferredMeas}).
It is also straightforward to apply the techniques of this section to simulate classically controlled Clifford operations, provided the values of the controls do not arise from measurement outcomes.

Unitary stabilizer circuits admit a simple representation in terms of linear transformations modulo $D$, simplifying the individual operations used in the existing formalisms for $d = 2$ and $d \ge 2$ (in Refs.~\cite{GottPhD} and~\cite{HDM05}).
We use this simplified formalism to develop a more general stabilizer formalism which can describe the evolution of states under measurement, for arbitrary qudit dimension $d$, in Section~\ref{sec:fullStabilizerFormalism}.

\subsection{Stabilizer tableaus for qudits}
\label{sec:stabilizerTableaus}

We begin with a substitute for the so-called binary representation~\cite{DM03,AG04} of Pauli operators:

\begin{definition}
  A \emph{Pauli vector} on $n$ qudits is a vector
 $\bar{\vec v} \in \Z_D^{2n+1}$ (which we represent as a column vector to enable transformations by left-multiplication), which we decompose into blocks $\bar{\vec v} = \mbox{$[\, \phi \;|\; v_1 \;\; v_2 \;\; \cdots \;\; v_{2n} \,]\trans$}$.
\end{definition}
\noindent 
Every Pauli operator $P$ with order $\le d$ is proportional to a Weyl operator by a power of $\tau^{2}$.
It suffices to show $P^d = (\tau^h W_{\vec v})^d = \idop$ if and only if $\tau^{dh} = 1$, and solve modulo $D$ for $h$.
Thus we may use Pauli vectors to represent any Pauli operator of order at most $d$, using the correspondence
\begin{align}
	\label{eqn:pauliVectorCorresp}
	\mbox{\small
	$\left[\begin{array}{c}
			\!\phi\!
		\\[0.2ex] \\[-3ex]\hline \\[-3ex]
			\!\vec v\!
		\\[0.2ex]
	\end{array}\right]$} 
	\;\longmapsto\;	\tau^{-2\phi} \;\! W_{\vec v}.
\end{align}
As in the binary case, a \emph{stabilizer group} is an abelian subgroup of Pauli operators which each have $+1$-eigenspaces.
(Such operators $P$ may be represented by Pauli vectors: we only have $P^d \ne \idop$ if $d$ is even and $P^d = -\idop$, in which case $P$ has no $+1$-eigenspace.)
Such a group has a joint $+1$-eigenspace: for a set $\sS = \ens{S_1, \ldots, S_\ell}$ of generators, the projector onto the $+1$-eigenspace of $S \in \sS$ is given by
\begin{align}
		\Pi_S	\;=\;	\frac{1}{d}\,	\sum_{j = 1}^d	S^j	;
\end{align}
by expanding $\Tr(\Pi_{S_1} \Pi_{S_2} \cdots \Pi_{S_\ell})$ as a sum of traces of Pauli operators, we find that the trace is non-zero.

\begin{definition}
 	A \emph{stabilizer tableau} $T_{\sS}$ on $n$ qudits is a matrix over $\Z_D$ with $2n+1$ rows, whose columns are Pauli vectors for the generators $\sS = \{ S_1, S_2, \ldots, S_\ell \}$ of a stabilizer group.
	The first row $\bm\phi_\sS$ of the tableau is called the \emph{phase vector}, and the rest $\sfW_\sS$ the \emph{Weyl block} of the tableau.
	We may present $T_\sS$ in terms of $n \x \ell$ blocks $\mathsf X_\sS$ and $\mathsf Z_\sS$\;,
	\begin{align}
		\label{eqn:stabilizerTableau}
		T_\sS
		\;=\;
			\left[\;\, \begin{matrix}	\bm\phi_\sS	\\	\\[-2.5ex] \hline \\	\mspace{20mu} \sfW_\sS \mspace{20mu} \\[3ex] \end{matrix}\;\,\right]
		\;=\;
			\left[\;\, \begin{matrix}	\bm\phi_\sS	\\	\\[-2.5ex]\hline \\[-2ex]	\mspace{20mu} \sfZ_\sS	\mspace{20mu} \\[1ex] \hline \\[-2ex] \sfX_\sS \end{matrix}\;\,\right].
	\end{align}
\end{definition}
\noindent
This definition corresponds closely to the notation in Refs.~\cite{GottPhD,AG04,DM03} for the binary case.
The principal distinction between this representation and the binary representation is that the $\mathsf Z_\sS$ and $\mathsf X_\sS$ blocks represent coefficients for Weyl operators, which differ from simple powers of $Z$ and $X$ operators on various qudits by scalar factors; and also that the coefficients are defined modulo~$D$, rather than modulo~$d$.

The more important distinction is between Weyl operators and simple powers of $Z$ and $X$: while subtle, this difference is significant because of the product formula for Weyl operators in Lemma~\ref{lemma:WeylCalculus}.
Consider stabilizer tableaus satisfying a further constraint:
\definition\
  A stabilizer tableau is \emph{proper} if the columns of its Weyl block $\sfW_\sS$ are orthogonal (mod~$D$) with respect to the symplectic inner product; \ie~if $\sfW_\sS\trans \sigma_{2n} \sfW_\sS \equiv 0 \pmod{D}$.
\enddefinition
\noindent
As the generators $S \in \sS$ of a stabilizer group commute, the columns of the Weyl block in a corresponding tableau are orthogonal mod~$d$ under the symplectic inner product by Lemma~\ref{lemma:WeylCalculus}; proper tableaus simply satisfy the same constraints mod~$D$.
Not all stabilizer groups can be represented by proper tableaus (see Section~\ref{sec:extendedTableaus} for a simple counterexample);\label{discn:notAlwaysProper} however, every standard basis state $\ket{q_1 \, q_2 \cdots \,q_n} \in \cH_d\sox{n}$ has a proper tableau of the form
\begin{align}
	\label{eqn:stdBasisTableau}
			\left[\;\; \begin{matrix}
				\mspace{-9mu}
					\begin{array}{c@{\;\;}c@{\;\;\cdots\;\;}c} q_1 & q_2 & q_n	\end{array}
				\mspace{-9mu}
			\\ \\[-2.5ex]	\hline \\[-2ex]
					\idop_n
			\\[1ex] \hline \\[-2ex]
					0_n
			\\[1ex]
			\end{matrix}\;\;\right]  .
\end{align}
We may show this by noting that 
$\ket{q_j} \in \cH_d$ is the state stabilized by $\tau^{-2q_j} Z$ acting on the $j$\textsuperscript{th} qudit~\footnote{
	Note that the $+1$-eigenstates of an operator $\tau^{-2\phi} W_{\vec v}$ can also be described as $\tau^{2\phi}$-eigenstates of $W_{\vec v}$.
	Phase coefficients may thus be used to denote powers of $\tau^2$ as eigenvalues, describing a stabilized space as an intersection of the corresponding eigenspaces of the Weyl operators.}.
Furthermore, every stabilizer group has a proper tableau for $d = 2$ or $d$ odd; in Appendix~\ref{apx:makeProper}, we show how to obtain such a tableau from an ``improper'' one.
We may thus restrict our attention to proper tableaus in many cases.

Given a proper stabilizer tableau, in which we represent Pauli operators in terms of Weyl operators, we may represent recombinations of stabilizer generators using only linear operations on columns, with no further computation of phases:
\lemma\
	For two Pauli vectors $\phi_1 \oplus \vec v_1$ and $\phi_2 \oplus \vec v_2$ drawn from a proper stabilizer tableau, we have
	\begin{align}
			\bigl[\tau^{-2\phi_1} W_{\vec v_1}\bigr] \bigl[\tau^{-2\phi_2} W_{\vec v_2} \bigr]
		=
			\tau^{-2(\phi_1 + \phi_2)} W_{\vec v_1 + \vec v_2}\,	.
	\end{align}
\endlemma\noindent
This follows directly from Lemma~\ref{lemma:WeylCalculus}.
Compare this to multiplication of bare products of $Z$ and $X$, in which ${(Z \ox Z)(X \ox X)} = {(ZX \ox ZX)}$, but ${(X \ox Z)(Z \ox X)} = {-(ZX \ox ZX)}$.
In the usual binary representation used \eg~in Ref.~\cite{AG04}, as well as the representation of Ref.~\cite{HDM05} for arbitrary $d \ge 2$, supplemental phase-factors must be computed depending on which generators are being multiplied; the phases depend quadratically on the exponents of $X$ and $Z$ involved in the operator.
Our formalism absorbs these quadratic dependencies into the calculus of Weyl operators, so that they do not appear elsewhere in computations involving products of generators.
Thus, changes in the generating set of a stabilizer group can be represented using only linear transformations on a proper tableau.
As we see in the next section, a representation using Weyl operators allows other transformations of the tableau to be effected linearly as well.

For $d$ even (\eg, for the case $d = 2$ corresponding to qubits), stabilizer tableaus are defined modulo $D = 2d$ rather than mod~$d$.
This difference from Refs.~\cite{GottPhD,AG04,DM03} is significant when using Weyl operators to represent Pauli operators, as illustrated in Eq.~\eqref{eqn:mod-d-negation}; and varying coefficients of the tableau by $\pm d$ may not preserve the property of the tableau being proper.
In Section~\ref{sec:extendedTableaus}, we describe a way in which the Weyl coefficients of a tableau may also be reduced mod~$d$  (though this involves techniques which accommodate improper tableaus).
However, the phase vector of a stabilizer tableau may always be evaluated modulo $d$, as differences of $d$ in phase coefficients amount only to a scalar factor of $\tau^{2d} = +1$ for the operator represented.

\subsection{The Clifford group on qudits}
\label{sec:genCliffordGroup}

\vspace*{-2.5ex}
\definition\
 	The \emph{Clifford group on $n$ qudits} is the group of operators $U \in \U(\cH_d\sox{n})$ such that $U P U\herm \in \cP_d\sox{n}$ for any $P \in \cP_d\sox{n}$; we denote it by $\Cliff_n(d)$.
\vspace*{-1.5ex}

\subsubsection{Linear transformations of stabilizer tableaus}

For any $U \in \Cliff_n(d)$, the superoperator $\Phi (M) := U M U\herm$ is an automorphism of matrix algebras;
we may then reduce the study of the Clifford group to its effect on a generating set $\ens{\tau\idop, Z_1, Z_2, \ldots, X_1, X_2, \ldots}$ of $\cP\sox{n}$.
Furthermore, $\Phi$ maps $\tau\idop$ to itself, and maps each Pauli operator of order $d$ to another Pauli operator of order~$d$.
Thus we can characterize the effect of an arbitrary Clifford operator on $n$ qudits by a $(2n+1)\x(2n+1)$ array over $\Z_D$ (a ``conjugation tableau''),
\vspace*{-6pt}
\begin{align}
	\label{eqn:ConjTableau}
	\sT_U
	=
  \mbox{\footnotesize$
	\left[\begin{array}{c|c@{\;\;}c@{\;\;}c}
		1 & h_1 &	\cdots &	h_{2n}
	\\
	\hline
		\begin{matrix} 0 \\[-0.75ex] \vdots \\[0.75ex] 0 \end{matrix} &			&		\text{\normalsize$C_U$} &	\\
	\end{array}\right]
	$},
\end{align}
whose columns are 
vectors representing the images of 
$\ens{\tau^2\idop, W_{\unit_1}, \ldots, W_{\unit_{2n}}}$ under $\Phi$, in sequence.
The effect of conjugating a set $\sS$ of operators given by a tableau as in Eq.~\eqref{eqn:stabilizerTableau} by a Clifford unitary $U$ can be evaluated by matrix multiplication modulo $\Z_D$\,:
\begin{align}
		\label{eqn:tableauTransform}
			\left[\;\, \begin{matrix}	\bm\phi_{U \sS U\herm}	\\[1ex]	\hline \\[-2ex]	\mspace{6mu} \sfW_{U \sS U\herm} \mspace{6mu} \\ \end{matrix}\;\,\right]
		\;=\;
			\sT_U
			\left[\;\, \begin{matrix}	\bm\phi_{\sS}	\\[1ex]	\hline \\[-2ex]	\mspace{6mu} \sfW_{\sS} \mspace{6mu} \\ \end{matrix}\;\,\right]
	.
\end{align}
We demonstrate this by identifying a class of Clifford operators whose conjugation tableaus have coefficients $h_j = 0$ for each $1 \le j \le 2n$:

\definition\
	A \emph{symplectic Clifford} operator $U \in \Cliff_n(d)$ 
	is one for which $U W_{\unit_j} U\herm$ is a Weyl operator for each $1 \le j \le 2n$.
	We write the set of such operators as $\sCliff_n(d)$.
\enddefinition

\noindent
For $U \in \C_n(d)$ with a tableau as in Eq.~\eqref{eqn:ConjTableau}, we may decompose $U = U'\; W_{\sigma_{2n} \vec h}$\,; by considering the effect of conjugating each of the Weyl operators $W_{\unit_j}$ by \smash{$U {W^{-1}}_{\mspace{-28mu} \sigma_{2n} \vec h}$}\,, we may show that $U' \in \sCliff_n(d)$.
We then prove Eq.~\eqref{eqn:tableauTransform} by relating the Weyl blocks $C_{U'}$ of $U' \in \sCliff_n(d)$ to symplectic transformations modulo $D$:

\vspace*{-0.6ex}
\definition\
	\label{def:symplClifford}%
	A \emph{symplectic transformation} of $\Z_D^{2n}$ is an operator $C : \Z_D^{2n} \to \Z_D^{2n}$ such that $C\trans\! \sigma_{2n} C  = \sigma_{2n}\mspace{2mu}$, so that $C$ preserves the symplectic inner product modulo $D$.
	We denote the group of such operators by $\Sp_{2n}(\Z_D)$.

\theorem\
	\label{thm:symplecticClifford}%
	For any $U \in \sCliff_n(d)$, there exists an operator $C \in \Sp_{2n}(\Z_D)$ such that
	\begin{align}
		\label{eqn:specialCliffordSymplecticCorresp}
		 	U W_{\vec v} U\herm
		\;=\;
			W_{C \vec v},
		\qquad
		\text{for all $\vec v \in \Z_D^{2n}$\,.}
	\end{align}
	In particular, $\sCliff_n(d)$ is a group.
	Similarly, for any $C \in \Sp_{2n}(\Z_D)$, there is a $U \in \sCliff_n(d)$ for which Eq.~\eqref{eqn:specialCliffordSymplecticCorresp} holds.

\proof\
	Any $C \in \Sp_{2n}(\Z_D)$ induces a superoperator $\Phi$ defined on linear operators acting on $\cH_d\sox{n}$, by extending $\Phi(W_{\vec v}) = W_{C\vec v}$ linearly~\footnote{%
		Note that in the case of $d$ even, the Weyl operators $W_{\vec v}$ for $\vec v \in \Z_D^{2n}$ are not linearly independent; then we must show that such a map $\Phi$ is well-defined.
		However, by the discussion following Lemma~\ref{lemma:weylOrthogonal}, $W_{C \vec v}$ cannot be expressed as a linear combination of other Weyl operators except if $W_{C \vec v} = \pm W_{C \vec w}$ for some vector $C\vec w \in \Z_D^{2n}$, in which case the sign is given by $(-1)^{\sympl{C \vec v, C \vec w}} = (-1)^{\sympl{\vec v, \vec w}}$ by Lemma~\ref{lemma:WeylModular}.
		Then $W_{\vec v} = \pm W_{\vec w}$ with the same sign, so that $\Phi$ as described above is indeed well-defined.}.
	As $C$ is invertible, so is $\Phi$; then $\Phi \ox \idop$ is a matrix automorphism (where $\idop$ is the identity superoperator on any auxiliary space).
	In particular, ${\Phi \ox \idop}$ maps projectors to projectors; thus it is positive, so that $\Phi$ itself is completely positive.
	Finally, as $\Phi(\idop) = \idop$ and $\Phi(W_{\vec v})$ has trace zero for $W_{\vec v} \ne \idop$, it follows that $\Phi$ is trace-preserving.
	Then $\Phi$ is an invertible CPTP map, so that $\Phi(M) = U M U\herm$ for some $U \in \sCliff_n(d)$.

  Conversely, for $U \in \sCliff(\cH_d\sox{n}$), the map $M \mapsto U M U\herm$ preserves group commutators $W_{\vec v} W_{\vec w} W_{\vec v}\herm W_{\vec w}\herm$ of Weyl operators within the Pauli group.
	This constrains the Weyl block $C_U$ of $\sT_U$: for any $1 \le h,j \le 2n$, if $UW_{\unit_h}U\herm = W_{\vec v_h}$ and $UW_{\unit_j}U\herm = W_{\vec v_j}$, we find that 
	$\sympl{\vec v_h, \vec v_j} \equiv \sympl{\unit_h, \unit_j} \!\pmod{d}$.
	Thus, the $2n \x 2n$ array $C_U$ in fact represents a linear transformation of $\Z^{2n}_D$ which is symplectic modulo $d$.
	For $d$ even, we then ``lift'' $C_U$ to obtain a matrix $C'_U \equiv C_U \!\pmod{d}$ which is symplectic mod $2d$.
	(The procedure to do so is technical, and similar to the procedure for producing proper stabilizer tableaus for $d$ prime or $d$ odd: we defer this detail to Appendix~\ref{apx:makeSymplectic}.)
\QED\endproof

\vspace*{-0.6ex}
\noindent
The above Theorem essentially extends \cite[Lemma~2]{Appleby05} to the multi-qudit case, albeit expressed here explicitly in terms of symplectic operators $C$ (rather than invertible operators, which is only equivalent in the case $n = 1$).

\vspace*{-0.6ex}
\corollary\
	The action of $U \in \Cliff_n(d)$ by conjugation on Pauli operators 
	may be characterized by Eq.~\eqref{eqn:tableauTransform}.

\vspace*{-0.6ex}
\proof\
	This is easy to show for Pauli operators, and follows for special Clifford operators from the characterization in terms of $\Sp_{2n}(\Z_D)$.
	Then $\sT_U$ is a product of linear operators arising from these special cases.
\QED\endproof

While Refs.~\cite{DM03,HDM05} also relate the Clifford group to symplectic operators, 
we have shown more specifically that $\sCliff_n(d) \subset \Cliff_n(d)$ is a group representation of $\Sp_{2n}(\Z_D)$.
As well, 
because stabilizer tableaus transform linearly under Clifford operators (with the action on the Weyl block of the tableau being symplectic), we obtain:

\corollary\
		Acting on a proper tableau $T_\sS$ by any conjugation tableau $\sT_U$ yields another proper tableau.

\subsubsection{Generators for the symplectic Clifford group}

The symplectic Clifford group not only has convenient algebraic properties, but includes the best-known Clifford operators on qudits from the folklore, and analogues of operators from generalizations of the Pauli group to finite fields (\eg~Ref.~\cite{GRB03}), including the single-qudit operators
\vspace*{-2ex}
\begin{subequations}%
\label{eqn:spExamples}%
\begin{equation}
\begin{gathered}
\begin{aligned}
			S			&=	\sum_{q = 0}^{d-1}	\tau^{q^2} \ket{q}\bra{q}	,
		&\;\,
			F 		&=	\frac{1}{\sqrt d} \sum_{p,q=0}^{d-1} \tau^{2pq} \ket{p}\bra{q},
\end{aligned}
\\
			M_a		=	\sum_{q=0}^{d-1} \ket{aq}\bra{q},
								~\text{for $a \in \Z_d^\ast$}\;,
\end{gathered}
\end{equation}%
where $\Z_d^\ast$ stands for the multiplicative group of units modulo $d$ and where multiplication is performed mod $d$; as well as both of the two-qudit operators
\begin{align}
			\cZ		\,&=	\sum_{q_1=0}^{d-1}\sum_{q_2 = 0}^{d-1}	\tau^{2q_1q_2} \ket{q_1}\bra{q_1} \ox \ket{q_2}\bra{q_2}	\,;
		\\[1.5ex]
%
			\cX		\,&=	\sum_{q_1=0}^{d-1}\sum_{q_2 = 0}^{d-1} 	\ket{q_1}\bra{q_1} \ox \ket{q_2 + q_1}\bra{q_2}\,.
\end{align}
\end{subequations}
In the binary case, $S$ is the $\pi/4${\:\!}-phase gate and $F$ is the Hadamard gate while $\cZ$ and $\cX$ are the controlled-$Z$ and \cnot\ gates.
For $d \ge 2$ in general, $F$ is the quantum Fourier transform; and the operators $M_a$\,, $X$, and $\cX$ can be used to perform any invertible affine transformation modulo $d$ on the computational basis.

For $U$ unitary, we write \smash{$A \mapstoname{ U} B$} to denote the relation $U \! A\;\! U\herm = B$ for the sake of brevity.
One may easily verify that the above operators $U$ are symplectic Clifford operators by computing the effect of $U W_{\unit_j} U\herm$ on standard basis states, for $1 \le j \le 2n$, to verify the following equations:
\begin{subequations}
\label{eqn:specialCliffordGentorsMap}
	\begin{align}
	\begin{split}
			Z &\mapstoname{S} 	Z ,
	\\[-1ex]
			Z &\mapstoname{F} X^{-1} ,
	\\[-1ex]
			Z &\mapstoname{M_a} Z^{\,a\!\!\:^{-\!\!\:1} \text{(mod $D$)}},
	\end{split}
	&
	\begin{split}
			X &\mapstoname{S}	W_{1,1}	\;,
	\\[-1ex]
			X &\mapstoname{F} Z	,
	\\[-1ex]
			X &\mapstoname{M_a} X^a	,
	\end{split}
\\[2.5ex]
	\begin{split}
	   Z \ox \idop &\mapstoname{\cZ} Z \ox \idop\,,
	\\[-0.5ex]
	   \idop \ox Z &\mapstoname{\cZ} \idop \ox Z\,,
	\end{split}
	&
	\begin{split}
	   X \ox \idop &\mapstoname{\cZ} X \ox Z\,,
	\\[-0.5ex]
	   \idop \ox X &\mapstoname{\cZ} Z \ox X\,,
	\end{split}
\\[2.5ex]
%
	\begin{split}
	   Z \ox \idop &\mapstoname{\cX} Z \ox \idop\,,
	\\[-0.5ex]
	   \idop \ox Z &\mapstoname{\cX} Z\herm  \ox Z\,,
	\end{split}
	&
	\begin{split}
	   X \ox \idop &\mapstoname{\cX} X \ox X\,,
	\\[-0.5ex]
	   \idop \ox X &\mapstoname{\cX} \idop \ox X\,,
	\end{split}
	\end{align}
	all of which are mappings from Weyl operators to Weyl operators~\footnote{
		In the case of $M_a$, we may represent $a \in \Z_D\units$ by an integer $0 < \alpha < D$ which is coprime to $d$; then $\alpha$ is also coprime to $D$, and the expression $\alpha^{-1}$ represents an integer for which $\alpha \alpha^{-1} \equiv 1 \pmod{D}$.
		There are at most two such integers $0 < \alpha < D$; it is easy to show that the operators $M_a$, $X^a$, and $Z^{\smash{a^{\text{--}1}}}$ arising from them will be the same.}.
	\end{subequations}%
	Thus, each such $U$ may be represented by conjugation tableaus of the form $1 \oplus C_U$, for the following symplectic operators $C_U \in \Sp_{2n}(\Z_D)$\,:
	\begin{align}
	\begin{split}
			C_S			&= \left[\mbox{\small $\,\begin{matrix} 1 \;&\; 1 \\ 0 \;&\; 1 \end{matrix}\,$}\right]\!,
	\\[2ex]
			C_F 		&= \left[\mbox{\small $\,\begin{matrix} \;\:0 \;&\; 1 \\ \!-1 \;&\; 0 \end{matrix}\,$}\right]\!,
	\\[2ex]
			C_{M_a} &= \left[\mbox{\small $\,\begin{matrix} a^{-1} & 0 \\ \!\!\!0\,\, & a \end{matrix}\,$}\right]\!,
	\end{split}
	&
	\begin{split}
			C_{\cZ} &= \left[\mbox{\small $\,\begin{matrix}
																1 \;&\; 0 \;&\; 0 \;&\; 1 	\\
																0	\;&\;	1	\;&\;	1	\;&\;	0		\\
																0	\;&\;	0	\;&\;	1	\;&\;	0		\\
																0	\;&\;	0	\;&\;	0	\;&\;	1 	\end{matrix}\,$}\right]\!,
	\\[1ex]
			C_{\cX} &= \left[\mbox{\small $\,\begin{matrix}
																1 \;&\!\! -1 \;&\; 0 \;&\; 0 	\\
																0	\;&\;	1	\;&\;	0	\;&\;	0		\\
																0	\;&\;	0	\;&\;	1	\;&\;	0		\\
																0	\;&\;	0	\;&\;	1	\;&\;	1 	\end{matrix}\,$}\right]\!.
	\end{split}
	\end{align}
Not only do these operators represent a collection of well-known operators which happen to belong to the symplectic Clifford group, they also characterize it:
\lemma
	\label{lemma:generateSymplClifford}
  The operators $S$, $F$, $\cZ$ (or $\cX$), and $M_a$ for $a$ ranging over the multiplicative units of $\Z_D$, generate $\sCliff_n(d)$ up to global phase factors.
\proof\
	We show this result by showing that the operators $C_{\cX}$, $C_S$, $C_F$, and $C_{M_a}$, together with direct sums with $\idop_2$ corresponding to tensor products with the identity operator on qudits~\footnote{
		In the construction of Ref.~\cite[Sec.~IV]{HDM05}, as in the formalism of this article, one- and two-qudit operations in many-qudit arrays are represented by applying a suitable choice of embedding of $\Sp_2(\Z_D)$ and $\Sp_4(\Z_D)$ into $\Sp_{2n}(\Z_D)$, specifically one which respects the indexing of the qudits being acted on.},
	generate the group $\Sp_{2n}(\Z_D)$.
	This result is technical, and we prove it by reduction to a similar result shown in Ref.~\cite[Sec.~IV]{HDM05} (see note~\footnote{
		Note that as we only use results of Hostens \etal~\cite{HDM05} concerning generation of $\Sp_{2n}(\Z_D)$ which hold for arbitrary moduli $D \ge 2$ and size $2n > 0$, the fact that our representation of stabilizer tableaus differs from theirs does not play any role in the proof.}).
	Using Eq.~\eqref{eqn:specialCliffordGentorsMap}, one may show that $\cX = (\idop \ox F\herm) \cZ (\idop \ox F)$, so that selecting either $\cX$ or $\cZ$ as a generator is equivalent.
	We may also describe $\cX_{2,1}$, a reversed version of $\cX_{1,2} := \cX$ (\ie~where $\cX_{2,1}$ has the second qudit the control, and the first qudit the target), by $\cX_{2,1} = (F\herm \ox \idop) \cZ (F \ox \idop)$.
	It is easy to show that for $g \in \Z$, the operator $\cX_{2,1}^{-g}$ has a conjugation tableau of the form $1 \oplus C^{\,-g}_{\chi}$\,, where
	\begin{gather}
		\label{eqn:symplOppCX}
	  			C^{\,-g}_{\chi} = \left[\mbox{\small $\,\begin{matrix}
																\,1 \;&\; 0 \;&\; 0 \;&\; 0 	\\
																\,g	\;&\;	1	\;&\;	0	\;&\;	0		\\
																\,0	\;&\;	0	\;&\;	1	\;&\!	-g		\\
																\,0	\;&\;	0	\;&\;	0	\;&\;	1 	\end{matrix}\,$}\right]\!.
	\end{gather}
	We can also generate $\textsc{swap}_{1,2}$ gates on pairs of qudits, which interchange $Z_1$ and $Z_2$, and similarly $X_1$ and $X_2$: routine calculation will show that we may decompose $\textsc{swap}_{1,2} = {(F^2 \ox \idop)} \cX_{1,2}\, \cX_{2,1}\herm\, \cX_{1,2}$\,. By construction, it has a conjugation map of the form $1 \oplus C_\varsigma$ for
	\begin{gather}
		\label{eqn:symplSwap}
	  			C_{\varsigma} = \left[\mbox{\small $\,\begin{matrix}
																\,0 \;&\; 1 \;&\; 0 \;&\; 0 	\\
																\,1	\;&\;	0	\;&\;	0	\;&\;	0		\\
																\,0	\;&\;	0	\;&\;	0	\;&\; 1		\\
																\,0	\;&\;	0	\;&\;	1	\;&\;	0 	\end{matrix}\,$}\right]\!.
	\end{gather}
	Finally, the conjugation map of the operator $F S^{-g} F\herm$ for $g \in \Z$ is the operator $1 \oplus C_F C_S^{\,-g} C_F^{-1}$, where one may compute
	\begin{gather}
		\label{eqn:symplOther}
	  			C_F C_S^{\,-g} C_F^{-1} = \left[\mbox{\small $\,\begin{matrix}
																\,1 \;&\; 0 \\
																\,g	\;&\;	1	\end{matrix}\,$}\right]\!.
	\end{gather}
	The three operators of Eqs.~\eqref{eqn:symplOppCX}~--~\eqref{eqn:symplOther}, together with the operators $C_F$ and $C_{M_r}$\,, are precisely those used in Ref.~\cite[Sec.~IV]{HDM05} to decompose arbitrary symplectic operators modulo $D$.

	Thus an arbitrary conjugation tableau for an operator $U \in \sCliff_n(d)$ may be generated by by those of the operators $S$, $F$, $\cZ$, and $M_a$\,; as $U$ is characterized up to scalar factors by its conjugation tableau, the Lemma follows.
\QED

\subsection{Deferring measurements in stabilizer circuits}
\label{sec:princDeferredMeas}

\PauliMeasurePreamble

For reasons that will become apparent in Section~\ref{sec:fullStabilizerFormalism}, describing the evolution of a state under a non-destructive measurement is more complicated for $d \ge 2$ arbitrary than for $d$ prime.
This is an obstacle to simulating circuits which involve both multiple measurements, and unitary operations which are conditioned on measurement outcomes.
We may overcome this by applying the principle of deferred measurement, postponing all measurements to the end without affecting the measurement statistics (or final residual quantum states).
This will allow us to efficiently simulate stabilizer circuits with classically-controlled Pauli operations (but not classically-controlled Clifford operations in general) using the techniques described thus far, so that we may consider the statistics of any single measurement unconditionally from any measurement which precedes it.
We may perform such simulations as follows.

\subsubsection{Reduction to $Z_r$ measurements}

As in the binary case~\cite{AG04}, there is no loss of generality in restricting from general Pauli measurements to ``non-destructive'' $Z$ measurements (\ie~in which the measured qudit has a defined post-measurement state).
For a $W_{\vec p} \in \cP_d\sox{n}$ measurement, this may be done in a straightforward way by constructing a $\cW_{\vec p}$ (``controlled $W_{\vec p}$'') operator, which on standard basis states $\ket{c} \in \cH_d$ and $\ket{\vec t} \in \cH_d\sox{n}$ would perform the operation
\begin{align}
		\label{eqn:controlledW}
		\cW_{\vec p} \ket{c} \ket{\vec t} \;=\; \ket {c} \ox W_{\vec p}^c \ket{\vec t}.
\end{align}
We may construct $W_{\vec p} \in \cP_d$ on a single qudit straightforwardly out of $\cZ$ and $\cX$ gates.
To illustrate this, consider the decomposition of $W_{\vec p}$ into $Z$, $X$, and a ``global phase'' gate, as illustrated in Fig.~\ref{fig:notionalCircuit}.
\par
\begin{figure}[h]
	\vspace*{2pt}
	\centerline{
\begin{picture}(175,28)(0,-12)
	\newcommand\gate[2][20]{\put(0,-8){\framebox(#1,16){\raisebox{-0.1ex}{\footnotesize $#2$}}}}

	\newcommand\wire[1][10]{\line(1,0){#1}}
	\def\puttext(#1,#2)#3{\put(#1,#2){\put(0,-3.5){#3}}}

	\put(0,0){
			\put(0,0){\wire[7]}
			\put(7,0){\gate[26]{W_{\:\!\!p_{\!\!\:1}\!\!\:,p_{\!\!\:2}}}}
			\put(33,0){\wire[7]}

		\puttext(50,0){$\equiv$}

		\put(70,0){\wire}
		\put(80,0){\gate{X^{p_2}}}
		\put(100,0){\wire[5]}
		\put(105,0){\gate{Z^{p_1}}}
		\put(125,0){\wire[5]}
		\put(130,0){\gate[35]{\tau^{-p_{\!\!\;1} \!\!\; p_{\!\!\;2}}\idop}}
		\put(165,0){\wire}
	}
\end{picture}  
}
	\vspace*{-8pt}
	\caption{\label{fig:notionalCircuit}%
		A notional circuit for a $W_{\vec p}$ gate on one qudit.}
\end{figure}
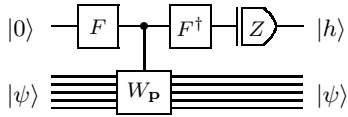
\begin{figure}[h]
	\vspace*{8pt}
	\centerline{
\begin{picture}(117,35)
	\newcommand\gate[2][20]{\put(0,-8){\framebox(#1,16){\raisebox{-0.1ex}{\footnotesize $#2$}}}}
	\newcommand\ctrl[1]{\put(0,0){\circle*{3}}\put(0,0){\line(0,-1){#1}}}
	\newcommand\wire[1][10]{\line(1,0){#1}}
	\def\puttext(#1,#2)#3{\put(#1,#2){\put(0,-3.5){#3}}}

	\put(-3,0){
	\puttext(3,25){$\ket 0$}
	\puttext(3,0){$\ket{\psi}$}
	
	\put(20,25){\wire}
	\put(30,25){\gate[15]{F}}
	\put(45,25){\wire[20]}
	\put(55,25){\ctrl{17}}
	\put(65,25){\gate[15]{F\herm}}
	\put(80,25){\wire}
	\put(90,25){
		\put(0,7){\line(0,-1){14}}
		\put(2,7){\line(0,-1){14}}
		\put(2,7){\line(1,0){8}}
		\put(2,-7){\line(1,0){8}}
		\puttext(4.5,0){\footnotesize $Z$}
		\qbezier(10,7)(14,4)(15,0)
		\qbezier(10,-7)(14,-4)(15,0)
	}
	\put(105,25){\wire}

	\multiput(20,6)(0,-3){5}{\wire[25]}
	\put(45,0){\gate{W_{\mathbf p}}}
	\multiput(65,6)(0,-3){5}{\wire[50]}

	\puttext(120,25){$\ket{h}$}
	\puttext(120,0){$\ket{\psi}$}
	}

\end{picture}      
} 
	\vspace*{8pt}
	\caption{\label{fig:measurementCircuit}%
		Circuit to perform a measurement of an $n$-qudit Weyl operator $W_{\vec p}$, using a $Z$ measurement on the top qudit to obtain the result.
		(This is illustrated here for measurement of a $\tau^{2h}$-eigenstate of $W_{\vec p}$, where the outcome will be $h$.)}
\end{figure}

\noindent
To obtain a coherently controlled version, given that $W_{\vec p}^c = W_{c\vec p} = \smash{\tau^{-c^2 p_1 \!\!\: p_2} Z^{cp_1} X^{cp_2}}$, we replace the $Z$ and $X$ gates with $\cZ$ and $\cX$ gates on a common control qudit, and act on the control with $S^{-p_1 p_2}$ satisfying $S^{-p_1 p_2} \ket{c} = \tau^{-c^2 p_1 p_2} \ket{c}$.
Then we may decompose $\cW_{\vec p} = {(S^{-p_1 p_2} \ox \idop) \cZ^{p_1} \cX^{p_2}}$ for single-qudit Weyl operators illustrated in Fig.~\ref{fig:controlledWCircuit}.
For a multi-qudit Weyl operator $W_{\vec p} = {W_{\!\!\:p_1, p_{n\!\!\;+\!\!\:1}} \ox W_{\!\!\:p_2, p_{n\!\!\;+\!\!\:2}} \ox{}} \cdots {\,\ox\, W_{p_n,p_{2n}}}$ 
where $n > 1$, we decompose $\cW_{\vec p}$ into operators $\cW_{p_j, p_{n+j}}$ acting with a common control and distinct targets.
\begin{figure}[h]
\vspace*{8pt}
	\centerline{
\begin{picture}(210,100)(0,-65)
	\newcommand\gate[2][20]{\put(0,-10){\framebox(#1,20){\footnotesize $#2$}}}
	\newcommand\ctrl[1]{\put(0,0){\circle*{3}}\put(0,0){\line(0,-1){#1}}}
	\newcommand\wire[1][10]{\line(1,0){#1}}
	\def\puttext(#1,#2)#3{\put(#1,#2){\put(0,-3.5){#3}}}

	\put(0,0){
		\renewcommand\gate[2][20]{\put(0,-8){\framebox(#1,16){\raisebox{-0.1ex}{\footnotesize $#2$}}}}
		\put(5,0){
			\put(0,0){\wire[7]}
			\put(7,0){\gate[26]{W_{\:\!\!p_{\!\!\:1}\!\!\:,p_{\!\!\:2}}}}
			\put(33,0){\wire[7]}
			\put(0,25){\wire[40]}
			\put(20,25){\ctrl{17}}
		}

		\puttext(50,12.5){$\equiv$}

	\put(-5,0){
		\put(70,0){\wire}
		\put(80,0){\gate{X^{p_2}}}
		\put(100,0){\wire[5]}
		\put(105,0){\gate{Z^{p_1}}}
		\put(125,0){\wire[45]}

		\put(70,25){\wire[60]}
		\put(90,25){\ctrl{17}}
		\put(115,25){\ctrl{17}}
		\put(130,25){\gate[30]{S^{-p_{\!\!\;1} \!\!\; p_{\!\!\;2}}}}
		\put(160,25){\wire}
	}
	}

	\put(0,-60){
		\put(5,0){
			\multiput(0,6)(0,-3){5}{\wire[10]}
			\put(10,0){\gate{W_{\mathbf p}}}
			\multiput(30,6)(0,-3){5}{\wire[10]}
			\put(0,25){\wire[40]}
			\put(20,25){\ctrl{15}}
		}

		\puttext(50,12.5){$\equiv$}

		\put(-5,0){
		\renewcommand\gate[2][20]{\put(0,-7){\framebox(#1,14){\raisebox{-0.1ex}{\footnotesize $#2$}}}}

		\put(70,10){\wire[6]}
		\put(76,10){\gate[33]{W_{\!p_{\;\!\!1}\!\!\:,p_{\!\!\;n\!\!\:{+}\!1}}}}
		\put(109,10){\wire[95]}

		\put(70,0){\wire[42]}
		\put(112,0){\gate[33]{W_{\!p_{\;\!\!2}\!\!\:,p_{\!\!\;n\!\!\:{+}\!2}}}}
		\put(145,0){\wire[60]}

		\put(150,-8){$\cdots$}

		\put(70,-10){\wire[96]}
		\put(166,-10){\gate[33]{W_{\!p_{\;\!\!n}\!\!\:,p_{2{\!\!\:}n}}}}
		\put(199,-10){\wire[6]}

		\put(70,25){\wire[77]}
		\put(150,22.5){$\cdots$}
		\put(164,25){\wire[41]}

		\put(92.5,25){\ctrl{8}}
		\put(127.5,25){\ctrl{18}}
		\put(182.5,25){\ctrl{28}}
	}}
\end{picture}    
} 
	\vspace*{8pt}
	\caption{\label{fig:controlledWCircuit}%
		Circuits for $\cW_{\vec p}$ gates, for $W_{\vec p} \in \cP_d$ and $W_{\vec p} \in \cP_d\sox{n}$.}
\end{figure}
Using the Fourier transform $F$ on the control qudit, we may apply standard techniques for eigenvalue estimation~\cite{K95} to perform a $W_{\vec p}$ measurement using a $Z$ observable measurement on the control, as illustrated in Figure~\ref{fig:measurementCircuit}.
\par 
\noindent
Measurement of a Pauli operator $P = \tau^{2\delta} W_{\vec p}$ may be performed similarly by replacing $\cW_{\vec p}$ with a controlled-$P$ gate, $\ctrlP = (Z^{\delta} \ox \idop\sox{n})\, \cW_{\vec p}$.
These substitutions may be easily performed at the time of simulation.

\subsubsection{Qudit states as measurement records}

While it is common to conceive of single-qudit measurements as being destructive, replacing the measured qudit with a classical record, we consider the qudit $r$ on which we perform the $Z$ observable measurement to be a persistent quantum system whose state is the measurement record.
To retain consistency with the notion of a classical measurement record, we may impose constraints of the following sort:
\begin{itemize}
\item 
	We may require that no operators act on the qudit $r$ (after the measurement) aside from those which commute with $Z_r$\,.
	For instance, it may act as the control of a product of $\cX$ or $\cZ$ operators, instead of acting as a classical control for Pauli operators on the same target qudits.
	In this case, the state of the qudit $r$ remains an unchanging record of the measurement outcome.
\item
	We may instead allow measured qudits $r$ to be subject to Pauli operators themselves, with classical control limited only to qudits $s$ which are also measurement records or input parameters.
	Such transformations of the measured qubits are limited to invertible transformations of the standard basis over $\Z_d$, and may be simulated in the stabilizer formalism by representing them as Clifford group operators.
\end{itemize}
Using these restrictions, we may commute $Z$ measurements on measurement registers past any Pauli operators which are conditioned on the outcome of that measurement.
If a measurement record acts only as a classical control for Pauli operations, both the distribution of outcomes for $r$ and the final state of any system depending on $r$ is unaffected by commuting the $Z_r$ measurement past the controlled operation.
More generally, while a controlled transformation depending on a measurement record $s$ will transform $Z_r$ non-trivially, the \emph{group} of observables $\gen{Z_m, \ldots, Z_r, Z_s}$ on the measurement records is preserved; so measuring $Z_m$, \ldots, $Z_s$ after the transformations will yield the same distribution of outcomes, as measuring $Z_m, \ldots, Z_s$ first and then transforming the outcomes.
With either approach, one may simulate all classically controlled operations $P \in \cP_d\sox{n}$ which depend on measurement outcomes using the corresponding \emph{coherently} controlled operators $\ctrlP$.
The distribution of any single measurement outcome may then be simulated using the techniques described thus far.
\subsubsection{Remarks on classically controlled Clifford operations}

Deferring measurements cannot be done for measurement records which control Clifford operations while remaining in the stabilizer formalism, using the techniques described above, except for controlled-Pauli operations.
Even in the simplest case, acting on states in which the control is in a standard basis state, such a simulation would involve transforming some columns of the tableau (\ie~representing Pauli operators supported on the target qudits of the operation) with different linear transformations depending on the values of other columns (which represent Pauli operators supported on the control qudit of the operation).
Such a transformation is at best multilinear, but not linear.
Furthermore, this transformation would be ill-defined for any column in which the control qudit is acted on by an operator with a non-zero $X$ component.
For instance, the coherently controlled operator $\cOp{S}$ obtained by deferring the measurement on the control acts on the operator $X \ox \idop$ as follows:
\begin{align}
		\lefteqn{\cOp{S} (X \ox \idop) \cOp{S}\herm \ket{c}\ket{t}}\quad
	\notag\\&=\;
		\left\{\begin{array}{@{\;}r@{~~}ll}
							(X \ox S) 			\ket{c}\ket{t},	&	\text{if $c \in \{0,\ldots,d-2\}$}	\,,	\\[1ex]
							(X \ox S^{1-d})	\ket{c}\ket{t},	&	\text{if $c = d-1$}	\,.		         
		       \end{array}\right.\!\!
\end{align}
Thus, $\cOp{S}$ fails to map $X \ox \idop$ to another Pauli operator via conjugation, and isn't represented by a transformation of stabilizer tableaus.
(Indeed, as it fails to preserve the Clifford group, such an operator $\cOp{S}$ is not itself a Clifford operator; supplementing the Clifford group with any single non-Clifford unitary generates a set of operators approximately universal for quantum computation for $d = 2$~\cite{NRS01} as well as any other prime dimension~\cite[Appendix~D]{CAB12}, so that classical algorithms to simulate \textsc{toffoli} gates together with Clifford operations are impossible unless $\BPP = \BQP$.)
To simulate a Clifford operation controlled by measurement outcomes, we must therefore explicitly compute (the probability distribution of) the measurement outcome, and use this outcome to determine mid-simulation which linear transformation to apply to the target qudits.

For any fixed measurement outcome, or for a classical control value provided as input (or generated according to a fixed probability distribution), simulating a classically controlled Clifford operation is no more difficult than a fixed Clifford operation; the only obstacle in this case is simulating the evolution of the state caused by the measurement itself, which we treat in Section~\ref{sec:fullStabilizerFormalism}.

\subsection{Computing the distribution of a single terminal measurement}
\label{sec:simTerminalMeas}

Describing the evolution of a state under measurement proves to be more technically complicated in the case of composite dimensions $d \geqslant 2$, as we shall see in Section~\ref{sec:fullStabilizerFormalism}.
However, determining the distribution of outcomes for a single measurement, on a state which can be represented by a proper stabilizer tableau, can be treated using the techniques described thus far.
We now summarize how to determine the distribution over possible outcomes, deferring the account of how states transform under measurement for Section~\ref{sec:fullStabilizerFormalism}, and proofs of certain details to Appendix~\ref{apx:spectraWeyl}.

Consider a proper stabilizer tableau $T_\sS$ which represents a group $G_\sS$ stabilizing a unique state.
To simulate measurement with respect to a Pauli operator $P$, consider some Weyl operator $W_{\vec p} \propto P$.
While $W_{\vec p}$ may not commute with all elements of $G_\sS$, there may be a non-trivial power of $W_{\vec p}$ which does.
The smallest positive power $s \ge 1$ such that $W_{\vec p}^{s}$ does commute with all of $G_\sS$ constrains the distribution of possible measurement outcomes for $P$.
By Lemma~\ref{lemma:WeylCalculus}, $s$ is the smallest positive integer such that $[s \vec p, \vec v_k] \equiv 0 \pmod{d}$, where $\varphi_k \oplus \vec v_k$ is the $k\textsuperscript{th}$ column of $T_\sS$.
That is, $s$ is the smallest positive integer such that
\begin{align}
		\label{eqn:terminalMeasurementCommutationCongruence}
		s (0 \oplus \vec p)\!\trans (0 \oplus \sigma_{2n}) \,T_\sS \;\equiv\; \vec 0\trans \pmod{d}	\;.
\end{align}
From the minimality of $s$, it follows that $s$ will be a factor of $d$, with $s = 1$ in the case that $W_{\vec p}$ commutes with $G_{\sS}$.
Specifically, let $\bm \phi = {\bigl(0 \oplus \vec p\bigr)\!\trans \bigl(0 \oplus \sigma_{2n} \bigr) \, T_\sS}$, and consider the greatest common factor $\eta$ of the coefficients of $\bm\phi$ together with $d$: then $s = d/\eta$.
As $W_{\vec p}^{s}$ commutes with all of the generators of $G_\sS$, and as $G_\sS$ stabilizes a unique state, we may show (Lemma~\ref{lemma:maximalStabilizer}) 
that $W_{\vec p}^{s}$ is proportional to an element of $G_\sS$.
Then the pre-measurement state is an eigenstate of $W_{\vec p}^{s}$, whose eigenvalue constrains the possible measurement outcomes.
If it is a $\tau^{2t}$-eigenstate, then $\tau^{-2t} W_{s\vec p}$ stabilizes the state, in which case there is a vector $t' \oplus s\vec p'$ in the column-span of $T_\sS$ such that $\tau^{-2t} W_{s\vec p} = \tau^{-2t'} W_{s\vec p'}$.
In order for this equality to hold, we require that $s\vec p' - s\vec p \equiv 0 \pmod{d}$, and that $2t' - 2t \equiv \sympl{s\vec p',s\vec p} \pmod{D}$ as well.
\begin{itemize}
\item
	For $d$ odd, these two conditions imply that $t' - t \equiv 0 \pmod{d}$, so that we simply need to determine for which value of $t \in \Z_D$ that $t \oplus s\vec p$ is in the column span of $T_\sS$.

\item
	For $d$ even, the vector $t' \oplus s\vec p'$ may differ from $t \oplus s\vec p$.
	In particular, we may have $t'$ and $t$ may differ by $\frac{d}{2}$ modulo $d$; and the vector $d \vec x = s\vec p' - s\vec p$ corresponds to some representation of the identity operator $\idop = W_{d\vec x}$ which accounts for the difference in the phase coefficients, by the formula
	\begin{equation}
		\begin{aligned}[b]
				\tau^{-2t'} W_{\!s\vec p'} \;&=\;	\tau^{-2t + \sympl{d\vec x,s\vec p}} \,W_{\!s\vec p + d\vec x} \\&=\; \tau^{-2t} \,W_{\!s\vec p} \,W_{\!d\vec x}	\;,		  
		\end{aligned}
	\end{equation}
	by Lemma~\ref{lemma:WeylModular}.
	We may attempt to account for the contribution of $W_{d\vec x}$ to the phases by introducing auxiliary columns of the form $\vec u_j := \frac{d}{2}\sympl{\unit_j, s\vec p} \oplus d\unit_j$ to the tableau $T_\sS$: adding $\vec u_j$ to $(t \oplus \vec p)$ yields another Pauli vector representing the same operator.
	We may also introduce a column $\vec u_0 = (d \oplus \vec 0)$, also representing the identity operator, to account for differences of $\pm d$ in the phase coefficient which are insignificant.
	Thus, it suffices to determine for which value of $t \in \Z_D$ that $(t \oplus s\vec p)$ is in the column span of \mbox{$\bigl[ \vec u_0 \; \vec u_1 \; \cdots \; \vec u_{2n} \;\big|\; T_\sS \;\bigr]$}.
\end{itemize}
In either case, it suffices to adjoin the vector $-(0 \oplus s\vec p)$ to the tableau (together with the vectors $\vec u_j$ in the case of $d$ even), and perform column-reductions to clear the Weyl tableau.
The resulting matrix will contain a vector $(t \oplus \vec 0)$, which implies that $(t \oplus s\vec p)$ is the the column span of the original matrix and that the pre-measurement state is a $\tau^{2t}$-eigenstate of $W_{\vec p}^s$.

The constraint that this imposes on the outcomes of a $W_{\vec p}$ measurement are as follows.
As $(\tau^{-2t} W_{\vec p}^{s})^\eta$ is proportional to $\idop$ and has $+1$-eigenstates, we have $2t \eta \equiv 0 \pmod{D}$, from which it follows that $t$ is a multiple of $s$.
The post-measurement state is a $\tau^{2u}$-eigenstate of $W_{\vec p}$ by definition for some $u \in \Z_d$, and in particular must be a $\tau^{2t}$-eigenstate of $W_{\vec p}^s$ as noted above; thus $2t \equiv 2su \pmod{D}$, or equivalently $u \equiv t/s \pmod{\eta}$.
We show in Appendix~\ref{apx:distributionMeasOutcomes} that the outcomes are in fact uniformly distributed over all $u \in \Z_d$ satisfying this constraint.
This characterizes the probability distribution over the outcome $\tau$ of the measurement of $W_{\vec p}$.
For $P = \tau^{-2\delta} W_{\vec p}$, the distribution of outcomes for a $P$ measurement are uniform over the solutions $h \in \Z_d$ to the congruence
\begin{align}
	\label{eqn:measOutcomeCongruence}
	h \;\equiv\; u + \delta \;\equiv\; t\eta/ d + \delta \pmod{\eta},
\end{align}
where again $\eta = \gcd(d,\phi_1,\phi_2,\ldots)$.
In particular, the results are uniformly distributed over some coset $\kappa + \eta \Z_d$, where we may take $\kappa = t\eta/d + \delta$.

Note that the above analysis accommodates the possibility of a deterministic outcome.
If $P$ commutes with $G_\sS$, then $\eta = d$, in which case Eq.~\eqref{eqn:measOutcomeCongruence} has a unique solution mod $d$: the outcomes are ``uniformly'' distributed over a singleton set (\ie~follow a delta-distribution).
This analysis also generalizes the familiar scenario of a random measurement outcome for $d$ prime: in that special case, we have $\eta = 1$ if the measurement observable $P$ does not commute with all elements of $\sS$.
In the case $\eta = 1$, Eq.~\eqref{eqn:measOutcomeCongruence} trivializes and imposes no constraints on $h \in \Z_d$, and so is uniformly distributed over $\Z_d$.

\subsection{Summary for simulating unitary stabilizer circuits}

We have shown that using Weyl operators to describe stabilizer tableaus allows us to describe a simple, linear formalism for representing the action of Clifford circuits on standard basis states, and determining the measurement outcomes for single measurements.
These results follow from the special algebraic properties of Weyl operators as described in Lemma~\ref{lemma:WeylCalculus}, and the fact that standard basis states may be represented by proper tableaus (as defined in Section~\ref{sec:stabilizerTableaus}).
Demonstrating a subgroup $\sCliff_n(d) \subset \Cliff_n(d)$ which is a representation of $\Sp_{2n}(\Z_D)$ is an additional dividend of this formalism.

The remainder of the article is devoted to extending this formalism to explicitly describe the evolution of stabilizer states under measurements, which involves providing a linear analysis of those states which lack ``proper'' stabilizer tableaus as described in Section~\ref{sec:stabilizerTableaus}.

\section{Improper tableaus and evolution under measurements}
\label{sec:fullStabilizerFormalism}

As we note on page~\pageref{discn:notAlwaysProper} (and show below), there do not always exist proper stabilizer tableaus for a stabilizer state in the case of even $d > 2$.
Furthermore, measurements may transform a state stabilized by a group with a proper tableau, to one stabilized by a group without any proper tableau.
This presents an obstacle for a uniform treatment of stabilizer circuits in the manner we have described so far, and are unavoidable when simulating the distributions of outcomes of circuits involving multiple measurements for even $d > 2$.
We must therefore extend the formalism described thus far to describe a linear formalism to simulate arbitrary stabilizer circuits, and simulate outcomes of multiple measurements.

\subsection{Extension to accommodate states having no proper tableaus}
\label{sec:extendedTableaus}

For even dimensions $d > 2$, there are stabilizer states whose stabilizer groups cannot be represented by a proper stabilizer tableau.
The simplest example is 
the single-qudit state $\frac{1}{\sqrt 2} \bigl( \ket{0} + \ket{2} \bigr)$ for $d = 4$, which is stabilized by the set $\sS = \ens{Z^2, X^2}$.
This set of generators has four possible stabilizer tableaus (up to column swaps and inconsequential changes in the phase block):
\begin{align}
  \mbox{\footnotesize$\left[\begin{array}{cc@{\;\;}c@{\;\;}c}~\\[-4ex]
			\!\ast & \ast\!
	\\\hline\\[-3ex]
			\!2 & 0\!
	\\
			\!0 & 2\!
	\end{array}\right]$},
&&
  \mbox{\footnotesize$\left[\begin{array}{cc@{\;\;}c@{\;\;}c}~\\[-4ex]
			\!\ast & \ast\!
	\\\hline\\[-3ex]
			\!2 & 0\!
	\\
			\!0 & 6\!
	\end{array}\right]$},
&&
  \mbox{\footnotesize$\left[\begin{array}{cc@{\;\;}c@{\;\;}c}~\\[-4ex]
			\!\ast & \ast\!
	\\\hline\\[-3ex]
			\!6 & 0\!
	\\
			\!0 & 2\!
	\end{array}\right]$},
&&
  \mbox{\footnotesize$\left[\begin{array}{cc@{\;\;}c@{\;\;}c}~\\[-4ex]
			\!\ast & \ast\!
	\\\hline\\[-3ex]
			\!6 & 0\!
	\\
			\!0 & 6\!
	\end{array}\right]$};
\end{align}
where each $\ast$ may independently be either $0$ or $4$ (representing a global phase of $+1$).
The reader may verify that none of these are proper, as the columns of the Weyl blocks are not orthogonal under the symplectic product modulo $D = 8$.
Thus, there do not exist any proper stabilizer tableaus for $\sS$; by linearity, there are no proper tableaus for any other set of operators generating the same group either.

As the Pauli vectors of commuting operators are always orthogonal mod $d$ under the symplectic inner product, such obstacles do not arise for $d$ odd.
In the case of $d = 2$, while ``improper'' tableaus do exist, every stabilizer state has a proper tableau, which may be obtained from an improper one by the techniques of Appendix~\ref{apx:makeProper}.
However, to describe a formalism which functions for arbitrary $d$, it is necessary to incorporate corrections to the phases as in the standard ``binary'' formalism, in order to describe different generating sets of the stabilizer group.
We show that these corrections may be subsumed in a more general linear formalism, as follows.

\subsubsection{Extended stabilizer tableaus}
\label{sec:defExtendedTableaus}

Consider a set of commuting Pauli vectors $\vec p_1, \ldots	, \vec p_\ell$, where each vector $\vec p_j = [\, \phi_j \,|\, \vec v_j \,]\trans$ represents a Pauli operator $P_j = \tau^{-2\phi_j} W_{\vec v_j}$.
By Lemma~\ref{lemma:WeylCalculus}, we have $W_{\vec u} W_{\vec v} = \tau^{\sympl{\vec u, \vec v}} W_{\vec u + \vec v}$\,, where $\sympl{\vec u, \vec v} \equiv \sympl{\vec v, \vec u}$ as $P_h$ and $P_j$ commute.
It follows that $\sympl{\vec u, \vec v} \in \ens{0,d}$ modulo $D$.
%
These represent corrections to the phase coefficients of Pauli vectors by multiples of $\frac{d}{2}$ (which would always be $0$ in the case of $d$ odd).
When computing products of the Pauli operators $P_j$, we may avoid computing the quadratic dependency arising from the symplectic inner products $\sympl{\vec v_h, \vec v_j}$ by storing an array $\Xi\in \{0,\frac{d}{2}, d, \frac{3d}{2}\}^{\ell \x \ell}$ such that $\Xi_{h,j} \equiv \frac{1}{2}\sympl{\vec v_h, \vec v_j} \pmod{D}$.
We may do this using an expanded table of coefficients, of the form
\begin{align}
		\bar T_\sS \,&= 
	\left[\begin{array}{ccc}
		\phi_1 & \cdots & \phi_\ell
	\\[0.2ex]
	\hline~
	\\[-1.5ex]
		\vec v_1 & \cdots & \vec v_\ell
	\\[1.5ex]
	\hline
	\\[-1ex]
		& \Xi &
	\\[-1ex]~
	\end{array}\right].
\end{align}
We call such an array an \emph{extended stabilizer tableau}, and refer to $\Xi$ as the \emph{phase correction block} of the tableau; the tableau $\bar T_\sS$ is proper if $\Xi \equiv 0 \pmod{D}$.
In the case of $d$ odd, this in fact holds by necessity.

Products and recombinations of generators may be represented by linear \emph{superoperators} on such extended tableaus, differing only slightly from left- and right-matrix-multiplications on $\bar T_\sS$.
For each $1 \le j \le \ell$, let $\bar{\vec p}_j = \vec p_j \oplus \Xi \unit_j$ be the column of $\bar T_\sS$ extending $\vec p_j$.
Consider a product of Pauli operators:
\begin{align}
	\tilde P \;=\; P_h P_j
	\;&=\;
		\tau^{-2(\phi_h + \phi_j +\, \Xi_{h\!,j})} \, W_{\vec v_h + \vec v_j}
 \;.
\end{align}
Let $\tilde\varphi = \phi_h + \phi_j +\, \Xi_{h,j}$ for the sake of brevity.
When taking products of $\tilde P$ with other operators $P_k$, the quadratic dependency involved in the phases depends only on the Weyl operator $W_{\vec v_h + \vec v_j}$\,, to which $\tilde P$ is proportional.
Therefore we have
\begin{align}
		\tilde P P_k
	\;&=\;
		\tau^{-2(\tilde\varphi \;\!+\;\! \phi_k \;\!+\;\! \sympl{\vec v_h \!\!\:+ \vec v_j, \vec v_k}/2)}
		\, W_{(\vec v_h + \vec v_j) + \vec v_k}
	\notag\\
	\;&=\;
		\tau^{-2(\tilde\varphi \;\!+\;\! \phi_k \;\!+\;\! [\Xi_{h,k} \,+\, \Xi_{j,k}])}
		\, W_{(\vec v_h + \vec v_j) + \vec v_k}	\,;
\end{align}
where the phase correction $\bigl[ \Xi_{h,k} + \Xi_{j,k} \bigr]$ is a sum of the phase corrections arising from $P_h$ and $P_j$.
Then, we may represent operator $\tilde P$ by an extended column vector of the form
\begin{align}
		\tilde{\vec p}
	\,&=
	\mbox{\footnotesize$
	\left[\;\begin{matrix}
		\!\!\phi_h + \phi_j + \Xi_{h,j}\!\!
	\\[0.2ex]
	\hline~
	\\[-2ex]
		\!\vec v_h + \vec v_j\!
	\\[1ex]
	\hline~
	\\[-2ex]
		\!\Xi\unit_h + \Xi\unit_j\!
	\\[-2ex]~
	\end{matrix}\;\right]
	$}
	\;=\;
	\bar{\vec p}_h + \bar{\vec p}_j +
	\mbox{\footnotesize$
	\left[\;\begin{matrix}
		\!\!\Xi_{h,j}\!\!
	\\[0.2ex]
	\hline~
	\\[-2ex]
		\vec 0
	\\[1ex]
	\hline~
	\\[-2ex]
		\vec 0
	\\[-2ex]~
	\end{matrix}\;\right]
	$}.
\end{align}
Thus to obtain the extended tableau $\bar T_{\tilde \sS}$ in which the generator $P_j$ is replaced with $\tilde P = P_h P_j$, it suffices to add the $h\textsuperscript{th}$ column into the $j\textsuperscript{th}$ column, subtract the $h\textsuperscript{th}$ row (of the phase-correction block) from the $j\textsuperscript{th}$ row to maintain the antisymmetry of that submatrix, and finally add the corrective term $\Xi_{h,j}$ into the phase coefficient.
That is, we compute
\begin{subequations}
\begin{align}
	\label{eqn:extendedTableauRecombine}
		\bar T_{\tilde \sS}
	\,&=\,
		\Psi_{h,j}\Bigl( E_{2n+\!\!\:1\!\!\:+\!\!\;h,\!\;2n+\!\!\:1\!\!\:+\!\!\;j}^{-1} \, \bar T_\sS \, E_{h,j} \Bigr),
\end{align}
	where $E_{a,b} = \idop \,+\, \unit_b \unit_a\trans$ is the usual elementary matrix for adding row $a$ into row $b$, and where
\begin{align}
		\Psi_{h,j}(M)
	\,&=\,
		M \:\!+\, \Bigl[ \unit_1\unit_{2n+1+h}\trans \Bigr] \:\! M \:\! \Bigl[ \unit_j\unit_j\trans \Bigr]
\end{align}
\end{subequations}
is a superoperator which adds the appropriate phase correction (stored in the $j\textsuperscript{th}$ column) for the operators represented by columns $h$ and $j$, to the phase coefficient of column $j$.

\subsubsection{Reducing the extended tableau modulo $d$}
\label{sec:tableauReductionMod-d}

In the case that $d$ is odd, we have $D = d$, and the entire tableau remains reduced modulo $d$ at all times.
However, if $d$ is even, coefficients $c \in \{d,d{+}1,\ldots,2d{-}1\}$ may arise.
Though these coefficients never present any actual difficulties in representing stabilizer states, it is still possible to reduce such coefficients modulo $d$ by performing phase corrections if this is desired.
As an extended tableau is under no restrictions to remain proper, we may more freely choose which Pauli vectors we use to represent a given Pauli operator generating a stabilizer group, so long as we maintain the correct phase correction information.

For a single Pauli vector, we may reduce the Weyl block modulo $d$ by considering different representations of the identity operator $\idop$ by Pauli vectors, such as
\begin{alignat}{6}
	\begin{split}
		\idop	\,&=	W_{d\unit_1}	=	Z_1^d,	\\
		\idop	\,&=	W_{d\unit_{n+1}}	=	X_1^d,	\\
	\end{split}
	&\quad
	\begin{split}
				&\ldots	\;, \\ &\ldots	\;,
	\end{split}
	&\quad
	\begin{split}
		\idop	\,&=	W_{d\unit_{n}}	=	Z_n^d,		\\
		\idop	\,&=	W_{d\unit_{2n}}	=	X_n^d	\;.
	\end{split}
\end{alignat}
This corresponds to Pauli vector representations of the form $0 \oplus d\unit_h$ for $1 \le h \le 2n$.
The phase correction required to combine such a vector with an arbitrary Pauli vector $\varphi \oplus \vec v$ is given by
\begin{align}
  \tilde\xi_h	\;=\;	\tfrac{1}{2}\sympl{d\unit_h, \vec v}	\;\equiv\;	\begin{cases}
																																\hfill \frac{d}{2} v_{h+n}	\;,	&	\text{if $1 \le h \le n$;}	\\[1ex]
																																-\frac{d}{2} v_{h-n}				\;,	&	\text{if $n < h \le 2n$,}
																															\end{cases}
\end{align}
modulo $D$.
Because $-\frac{d}{2} \equiv d + \frac{d}{2} \pmod{D}$ for $d$ even, and differences of $d$ have no effect in the phase columns, the phase corrections of $-\frac{d}{2}$ may be replaced by phase corrections of $+\frac{d}{2}$, or vice versa.
Reduction modulo $d$ of the Weyl coefficients of Pauli vectors can then be performed with the transformations
\begin{align}
	\begin{split}
		\varphi \oplus \vec v		\;&\longmapsto\;	(\varphi + \tfrac{d}{2}v_{n+h}) \oplus (\vec v - d\unit_h)	,	\\[1ex]
		\varphi \oplus \vec v		\;&\longmapsto\;	(\varphi + \tfrac{d}{2}v_{h}) \oplus (\vec v - d\unit_{n+h})	,
	\end{split}
\end{align}
as the vectors on the left and right all represent the same Pauli operator $\tau^{-2\varphi} W_{\vec v}$.
(This is used implicitly in the algorithm for determining distributions of measurement outcomes in Section~\ref{sec:simTerminalMeas}.)
By subsequently reducing the phase coefficient modulo $d$, the entire vector may be reduced modulo $d$.

While the above transformations do preserve which Pauli operator is represented by a given Pauli vector, it does not preserve the symplectic inner product of the Weyl block with those of other Pauli vectors.
To maintain an extended tableau while reducing Weyl coefficients modulo $d$, one must also update the phase correction block $\Xi$.
Suppose that $\varphi_j \oplus \vec v_j \oplus \Xi\unit_j$ is the $j\textsuperscript{th}$ column of an extended tableau $T_\sS$ which represents a set of operators $\{S_1, \ldots, S_j, \ldots, S_\ell\}$.
As with the operation to represent multiplication of generators described in Section~\ref{sec:defExtendedTableaus}, we may reduce the $h\textsuperscript{th}$ coefficient of $\vec v$ by $d$ by combining the phase correction information of the vector $0 \oplus d\unit_h$ for the other generators $S_1, \ldots, S_\ell$, to the vector $\vec \xi$ describing the phase correction information of $\varphi \oplus \vec v$.
That is, we compute the row vector
\begin{align}
	\label{eqn:reductionPhaseCorrection}
		\tilde{\vec \xi}_h
	\,&=\;
		\tfrac{1}{2}\Big[\;\begin{matrix}\sympl{d\unit_h, \vec v_1}	& \cdots & \sympl{d\unit_h, \vec v_\ell} \end{matrix}\;\Bigr]
	\notag\\&\equiv\;
		\tfrac{d}{2} \!\: \unit_h\trans \bigl[ \vec 0 \,\big|\, \sigma_{2n} \,\big|\, 0 \big] T_\sS \pmod{d},
\end{align}
and then add $\tilde{\vec \xi}_h$ to the $j\textsuperscript{th}$ row of the phase correction block $\Xi$ (updating the phase correction information for all other columns with respect to the $j\textsuperscript{th}$ column), and subtract $\smash{\tilde{\vec \xi}}_h\trans$ from $\vec \xi$ (updating the phase correction information for the $j\textsuperscript{th}$ column itself).
Note that by subtracting $\smash{\tilde{\vec \xi}}_h\trans$ from $\vec \xi$, the phase correction necessary for $\vec v$ is added into the $j\textsuperscript{th}$ coefficient of $\vec \xi$; before updating the $j\textsuperscript{th}$ row of the entire phase correction block, we may apply the superoperator $\Psi_{j,j}$ to perform the necessary phase correction in the $j\textsuperscript{th}$ column.
As the coefficients of the phase correction block are only ever added to the phase coefficients, they may themselves always be reduced modulo $d$ as in Eq.~\eqref{eqn:reductionPhaseCorrection}; thus whether the vector $\tilde{\vec \xi}_h$ and its transpose are added or subtracted is immaterial, except to maintain the antisymmetric property of the phase correction block.
\begin{subequations}
Thus, to reduce the $(h,j)$ coefficient of the Weyl block of an extended stabilizer tableau for $1 \le h \le n$ and obtain a new tableau $\bar T'_{\sS}$, we compute
\begin{equation}
\begin{aligned}[b]
				\bar T'_\sS
			\;&=\,
				\Psi_{j,j}\Bigl(T_\sS + \unit_{2n+1+j}\tilde{\vec \xi}_h \Bigr)
		\\&\quad\;\;- (0 \oplus \vec 0 \oplus \smash{\tilde{\vec \xi}}_h\trans)\unit_j\trans
				- d\unit_{1+h}\unit_j\trans
		\\[1ex]&=\;
				\Psi_{j,j}\Bigl(T_\sS + \tfrac{d}{2}\unit_{2n+1+j}\unit_{n+1+h}\trans T_\sS \Bigr)
		\\&\quad\;\;- \tfrac{d}{2}\bigl(0 \oplus \vec 0 \oplus [T_\sS\trans \unit_{n+1+h}]\bigr)\unit_j\trans
				- d\unit_{1+h}\unit_j\trans	;
\end{aligned}
\end{equation}
to reduce the $(h{+}n,j)$ coefficients of the Weyl block for $1 \le h \le n$, we instead compute
\begin{equation}
  \begin{aligned}[b]
				\bar T'_\sS
			\;&=\;
				\Psi_{j,j}\Bigl(T_\sS + \unit_{2n+1+j}\tilde{\vec \xi}_h \Bigr)
			\\&\quad\;\;- (0 \oplus \vec 0 \oplus \smash{\tilde{\vec \xi}}_h\trans)\unit_j\trans
				- d\unit_{n+1+h}\unit_j\trans
		\\[1ex]&=\;
				\Psi_{j,j}\Bigl(T_\sS + \tfrac{d}{2}\unit_{2n+1+j}\unit_{1+h}\trans T_\sS \Bigr)
			\\&\quad\;\;- \tfrac{d}{2}\bigl(0 \oplus \vec 0 \oplus [T_\sS\trans \unit_{1+h}]\bigr)\unit_j\trans
				- d\unit_{n+1+h}\unit_j\trans	.
  \end{aligned}
\end{equation}
In both cases, the mapping is an affine superoperator, involving both the superoperator $\Psi$ and transposition of the matrix $T_\sS$, and also of course the translation operation by $-d\unit_{1+h}\unit_j\trans$ or $-d\unit_{n+1+h}\unit_j\trans$ to reduce the actual coefficient of the tableau.
\end{subequations}

\subsubsection{Fixing the number of rows of extended stabilizer tableaus}	
\label{sec:extendedTableauSize}

When simulating unitary transformations on states represented by extended stabilizer tableaus, it may be preferable to use linear operators of a fixed size, independently of the state being transformed.
While the phase correction block is a square symmetric matrix, and may have a variable number of rows, we may bound its size if we can establish a maximum number of generators needed to describe an arbitrary stabilizer group.
We can then pad the number of rows or columns in any tableau on $n$ qudits to obtain tableaus with a fixed size.

As noted in Refs.~\cite{HDM05,Gheorg11}, a stabilizer group on $n$ qudits may have a minimal generating set strictly larger than $n$ when $d$ is composite.
The simplest example is the single-qudit state \mbox{$\frac{1}{\sqrt 2} ( \ket{0} + \ket{2} )$} described above for $d = 4$, for which $\smash{\{Z^2, X^2\}}$ is a minimal set of generators.
(Similar examples exist for any composite $d > 2$; one may consider a state stabilized by $\smash{\{Z^{d_1}, X^{d_2}\}}$ for arbitrary $d_1, d_2 > 1$ on qudits of dimension $d = d_1 d_2$.)
However, we may show that a minimal generating set for any $n$-qudit stabilizer group has at most $2n$ generators, as follows.
For any extended stabilizer tableau $\bar T_\sS$ (proper or otherwise) with $\ell > 2n$ columns, the Smith normal form $\mathsf S_\sS = L\, \sfW_\sS R$ of its Weyl block $\sfW_\sS$ has the form
\begin{align}
		\mathsf S_\sS
	\;=\;
		\left[\mbox{\small $\begin{array}{ccccc|ccc}  
													t_1			&	0				&	\cdots	&	\cdots				&	0				&	\\[-1ex]
													0				&	t_2			&					&								&	\vdots	&	\\[-0.5ex]
													\vdots	&					&	\ddots	&								&	\vdots	&	\; & \smash{\mbox{\Huge 0}} & \! \\[-1ex]
													\vdots	&					&					&	\!t_{2n-1}\!	&	0				&	\\[-0.5ex]
													0				& \cdots	&	\cdots	&	0							&	t_{2n}	&
		                    \end{array}$}\right]
\end{align}
where each coefficient $t_j$ is either zero or a divisor of $D$, and where $t_{j+1}$ is a multiple of $t_j$ for each $j < 2n$.
The right-most $(\ell - 2n)$ columns of $\sfW_\sS R$ are then also zero.
Interpreting $R$ as a recombination of the columns (and applying the appropriate phase corrections via Eq.~\eqref{eqn:extendedTableauRecombine}), we obtain an extended tableau $\bar T_{\tilde \sS}$ which has $\sfW_\sS R$ as a Weyl block.
The right-most $(\ell - 2n)$ columns then each represent the identity operator, and may be omitted to leave a tableau with at most $2n$ columns.
As the phase correction block requires only as many rows as columns, we may specify an arbitrary stabilizer group with an extended stabilizer tableau having at most $4n+1$ rows: one for the phase block, and $2n$ each for the Weyl and phase-correction blocks.
Thus, by padding, we may fix 
extended tableaus for $n$ qudit states to have $4n+1$ rows and $2n$ columns.

\subsection{Simulating unitary stabilizer circuits on extended tableaus}
\label{sec:extendedTableausClifford+TermMeas}

We may extend conjugation tableaus, representing the effect of conjugation of Pauli operators by Clifford operators, very simply.
As symplectic transformations do not affect the pair-wise symplectic inner products of the vectors $\vec v_j$, we may leave the phase correction blocks of the vectors $\bar{\vec p}_j$ untouched when we describe a transformation by $U \in \Cliff_n(d)$.
An extended conjugation tableau corresponding to a conjugation tableau $\sT_U$ can thus be given by a square matrix
\begin{align}
	\label{eqn:extenConjTableau}
	\mspace{-5mu}
	\bar\sT_U
	\,=
	\left[\begin{array}{c|c@{\;\;}c@{\;\;}c}
		\sT_U \;&\; 0
	\\[0.5ex]
	\hline~\\[-2ex]
		0 \;&\; \,\idop_{2n}\!
	\end{array}\right]
	=
  \mbox{\footnotesize$
	\left[\begin{array}{c|c@{\;\;}c@{\;\;}c|ccc}
		1 & h_1 &	\cdots &	h_{2n} & 0 & \cdots & 0
	\\
	\hline
		0 			& & 		& & \\[-1ex]
		\vdots	& & \smash{\mbox{\normalsize $C_U$}}	& & &	\smash{\mbox{\normalsize $0$}} \\
						& &     & & \\
	\hline
						& &     & &	\\[-1ex]
		\vdots	& & \smash{\mbox{\normalsize $0$}}		&	&	&	\smash{\mbox{\normalsize $\idop_{2n}$}}	\\[-0.5ex]
		0				&	&			&	&	
		\end{array}\right]
	$},
\end{align}
where again $C_U$ is a symplectic transformation of $\Z_D^{2n}$.

Simulating terminal measurements as in Section~\ref{sec:simTerminalMeas} requires only the ability to compute alternative generating sets for the stabilizer group as described in Eq.~\eqref{eqn:extendedTableauRecombine}, and the ability to compute symplectic inner products as in Eq.~\eqref{eqn:terminalMeasurementCommutationCongruence}.
For the latter, symplectic inner products of pairs of operators $\vec v_h, \vec v_j$ are not pertinent for computing the inner products $\sympl{\vec p, \vec v_h}$ and $\sympl{\vec p, \vec v_j}$ which are necessary to determine the statistics for measuring an observable $P = \tau^{-2\varphi} W_{\vec p}$. 
Thus it suffices to compute the smallest integer $s \ge 1$ such that
\begin{align}
 		\label{eqn:extended-terminalMeasurementCommutationCongruence}
		s \,\vec p\!\trans
		\left[\;\; \vec0 \;\;\Big|\;\, \sigma_{2n} \,\;\Big|\;\; 0 \;\;\right] \bar T_\sS
		\;\equiv\; \vec 0\trans \!\!\!\!\pmod{d}
\end{align}
as a simple modification of Eq.~\eqref{eqn:terminalMeasurementCommutationCongruence}.
We may then carry out a simple variation of analysis of Section~\ref{sec:simTerminalMeas}: defining 
$\smash{\tilde{\vec \xi}}_h\big.$ as in Eq.~\eqref{eqn:reductionPhaseCorrection}, we extend the tableau $\bar T_\sS$ with vectors $\tilde{\vec u}_h = 0 \oplus d\unit_h \oplus \smash{\tilde{\vec \xi}}_h\trans\big.$ for $1 \le j \le 2n$ in place of the vectors $\vec u_h$ and $\vec u'_h\big.$ described in Section~\ref{sec:simTerminalMeas}, and attempt 
to clear the Weyl block of the vector ${-(0 \oplus s \vec p \oplus \vec 0)}$, using Eq.~\eqref{eqn:extendedTableauRecombine} to perform column combinations.

Adopting these extensions to stabilizer tableaus and conjugation tableaus allow us to simulate arbitrary stabilizer states, whose stabilizer groups may or may not be representable by a proper tableau, using linear operations over $\Z_D$ as in Section~\ref{sec:simulateClifford}.

\subsection{Simulating stabilizer state evolution under Pauli measurements}
\label{sec:evolutionUnderMeas}

In many cases, including the case $d = 2$ and $d$ odd, every stabilizer state has a proper tableau, and simulating Pauli measurements in general may be achieved by transformations of proper tableaus.
However, for $d > 2$ even, such measurements may transform a state from one which can be represented by a proper stabilizer tableau, to one which cannot.
For instance, in the case $d = 4$, we may prepare the state 
\begin{align}
		\ket{\psi}	\;&=\; \cX_{a,b}^{\,2}\, F_a\:\! \ket{0}_a \ket{0}_b
	\notag\\[0.5ex]
		&=
	\;
			\tfrac{1}{2} \bigl( \ket{0} + \ket{2} \bigr)_a \!\ox\!\!\; \ket{0}_b  
			\;+\;
			\tfrac{1}{2} \bigl( \ket{1} + \ket{3} \bigr)_a \!\ox\!\!\; \ket{2}_b  ,
\end{align}
for which we may easily construct a proper tableau using the techniques of Section~\ref{sec:simulateClifford}.
However, if we measure $Z_b$ on $\ket{\psi}$, the post-measurement state on $a$ would differ from $\frac{1}{\sqrt 2} \bigl( \ket{0} + \ket{2} \bigr)$ by at most an $X$ operation.
As we saw in Section~\ref{sec:extendedTableaus}, such states have no proper stabilizer tableaus.
We must therefore extend the formalism of Section~\ref{sec:simulateClifford} to deal with evolution of states under measurement in the general case.

We now describe a procedure to describe this evolution for all $d \ge 2$, using the extended tableaus introduced in Section~\ref{sec:extendedTableaus}, deferring the proofs of certain statements to Appendix~\ref{apx:spectraWeyl}.
(In Section~\ref{sec:extendedSimplif}, we describe how this analysis may be simplified in the cases $d = 2$ and $d$ odd, for which extended tableaus are not in fact needed.)
Using the constructions of Sections~\ref{sec:princDeferredMeas} and~\ref{sec:extendedTableausClifford+TermMeas}, we may reduce the problem to that of describing the effect of a single-qudit $Z$ measurement.
(Again, this has the side-effect of allowing us to simulate classically controlled Pauli operations depending on the measurement result using a coherent controlled-Pauli operator.
We sketch a more direct approach to measurement in Section~\ref{sec:extendedSimplif}.)

Let $\sS = \{ S_1, \ldots, S_\ell \}$ be a minimal generating set for a stabilizer group $G_\sS$ which has a unique joint $+1$-eigenstate.
A $Z_r$ measurement has a non-trivial effect on a stabilizer state, and a non-deterministic outcome, if and only if the state is stabilized by a Pauli operator that does not commute with $Z_r$\,.
Given an extended stabilizer tableau $\bar T_\sS$\,, we may compute whether $Z_r$ commutes with each operator $S_k \in \sS$ by 
computing
\begin{align}
	\label{eqn:determineCommutingMeasurement}
		\bm \phi
	\,&=\,
		\unit_r\trans \! \left[\:\: \vec0 \:\:\Big|\,\, \sigma_{2n} \,\,\Big|\:\: 0 \:\:\right]  \bar T_\sS
	\,=\,
		(0 \oplus \unit_{n+r} \oplus \vec 0)\!\trans \bar T_\sS 
\end{align}
following the description in Eq.~\eqref{eqn:extended-terminalMeasurementCommutationCongruence}.
The result is a row-vector $\bm\phi = [ \phi_1 \; \ldots \; \phi_\ell ] \in \Z_D^{\,\ell}$ such that $Z_r S_k = \tau^{2\phi_k} S_k Z_r$.
As the row-vector $\bm\phi$ is the $(n+r+1)\textsuperscript{st}$ row of the tableau $T_\sS$, the tableau has the following block structure:
\begin{align}
	\mspace{-10mu}
	\bar T_\sS
	\,&=
  \mbox{\footnotesize$
	\left[\begin{array}{ccc@{\mspace{15mu}}l@{\mspace{-210mu}}}
		T_{1,1} &	\cdots &	T_{1,\ell} &	\gets \text{row $1$}
	\\
	\hline\\[-3ex]
		T_{2,1} 		& \cdots & T_{2,\ell} & \gets \text{row $2$}	\\ 
		\vdots			&					&	\vdots		&	\qquad\vdots	\\
		\phi_1			&	\cdots	&	\phi_\ell	&	\gets\text{row $n+r+1$}	\\
		\vdots			&					&	\vdots		&	\qquad\vdots	\\
		T_{2n+1,1}	&	\cdots	&	T_{2n+1,\ell}	&	\gets\text{row $2n+1$}	\\
	\hline
		 & 		& &  \\[-1ex]
		 & \smash{\mbox{\normalsize $\Xi_\sS$}} &	& \gets\text{phase correction block} \\
		 &     & & \\[-1ex]
		\end{array}\right]
	$}\mspace{150mu}
\end{align}
	We may recombine the generators described by the columns using Eq.~\eqref{eqn:extendedTableauRecombine}, to obtain a tableau $\bar T_{\tilde \sS}$ which has at most one column --- without loss of generality, the $\ell\textsuperscript{th}$ column --- containing an entry $\eta \not\equiv 0 \!\!\pmod{d}$ in the ${(n+r+1)\textsuperscript{st}}$ row:
	\begin{align}
	\label{eqn:columnReducedExtendedTableau}
	\mspace{-10mu}
	\bar T_{\tilde \sS}
	\,&=
  \mbox{\footnotesize$
	\left[\begin{array}{cccc@{\mspace{15mu}}l@{\mspace{-140mu}}}
		\tilde T_{1,1} 	&	\cdots	&	\tilde T_{1,\ell\!\!\;-\!\!\:1} & \tilde T_{1,\ell} &	\gets \text{row $1$}
	\\
	\hline~\\[-3ex]
		\tilde T_{2,1} 	& \cdots	&	\tilde T_{2,\ell\!\!\;-\!\!\:1} & \tilde T_{2,\ell} & \gets \text{row $2$}	\\ 
		\vdots					&					&	\vdots							&	\vdots				&	\qquad\vdots	\\
		0 							&	\cdots	&	0										&	\eta					&	\gets\text{row $n+r+1$}	\\
		\vdots					&					&	\vdots							&	\vdots				&	\qquad\vdots	\\
		\tilde T_{2n\!\!\:+\!1,1}	&	\cdots	&	\tilde T_{2n\!\!\:+\!1,\ell\!\!\;-\!\!\:1}	&	\tilde T_{2n\!\!\:+\!1,\ell}	&	\gets\text{row $2n+1$}	\\
	\hline
		 & 		& &  \\[-1.5ex]
		 & \smash{\mspace{40mu}\mbox{\normalsize $\Xi_{\tilde \sS}$}\mspace{-40mu}} &	& & \text{\qquad \vdots} \\
		 &     & & \\[-1ex]
		\end{array}\right]
	$}\mspace{105mu}
\end{align}
We may restrict $\eta$ to be a divisor of $d$ if it is non-zero, by multiplying the $\ell\textsuperscript{th}$ column by a suitable scalar $\alpha \in \Z_D^\ast$.
(To maintain the antisymmetry of the phase correction block, the $\ell\textsuperscript{th}$ row of $\Xi$ must also be multiplied by $\alpha$ in this case.)
Note that $\eta \in \{0,1,2\}$ in the case that $d = 2$, $\eta \in \{0,1\}$ in the case that $d$ is an odd prime, and $\eta \in \{0,d\}$ for any $d \ge 2$ in the case that $Z_r$ commutes with every generator of $\sS$.

The tableau $\bar T_{\tilde\sS}$ describes generators $\tilde \sS = \{\tilde S_1, \ldots, \tilde S_\ell\}$ for $G_\sS$ in which only $\tilde S_\ell$ may fail to commute with the observable $Z_r$\,, depending on the precise value of $\eta$.
Let $\tilde{\vec v}_j$ represent the Weyl block of the $j\textsuperscript{th}$ column of $\bar T_{\tilde \sS}$, and let $s = d/\eta$ (if $\eta > 0$), or $s = 1$ (if $\eta = 0$).
Then $s \ge 1$ is the smallest power of $\tilde S_\ell$ such that $\tilde S_\ell^s$ commutes with $Z_r$, as $\sympl{s\tilde{\vec v}_\ell, \unit_r} \equiv 0 \pmod{d}$ by construction.
Similarly, $Z_r^{s}$ commutes with $\tilde S_\ell$, and thus with every element of $\tilde \sS$.

\subsubsection{Computing the distribution of measurement outcomes}

As $G_\sS$ stabilizes a unique state, it follows (by Lemma~\ref{lemma:maximalStabilizer})
that $Z_r^{s}$ is proportional to some element of $G_\sS$.
We introduce the vector
\begin{align}
	\label{eqn:measurementObservablePhaseCorrectionVector}
		\vec \xi
	=
		\tfrac{1}{2}s\vec \phi\trans 
	=
		\begin{cases} 
									-\frac{d}{2} \unit_\ell	\,,	&	\text{if $d$ is even and $\eta > 0$;}
								\\
									\hfill \vec 0\,, \hfill & \text{if $d$ is odd or $\eta = 0$,}
		           \end{cases}
\end{align}
for the sake of brevity.
This vector then contains the phase correction information needed in order to represent products of $Z_r^{s}$, which we represent by the Pauli vector $(0 \oplus s\unit_r)$, with the generators $\tilde S_j$ in an extended tableau.
We find the element of $G_\sS$ to which $Z_r^{s}$ is proportional using a similar technique to Section~\ref{sec:extendedTableausClifford+TermMeas}, temporarily adjoining columns $\tilde{\vec u}_h = 0 \oplus d\unit_h \oplus \smash{\vec{\tilde{\xi}}}_h\trans$ for $\vec{\tilde{\xi}}_h$ as given in Eq.~\eqref{eqn:reductionPhaseCorrection} to aid in performing column reduction on the tableau, using Eq.~\eqref{eqn:extendedTableauRecombine} to perform column recombinations, to clear the Weyl block of the vector $-(0 \oplus s\unit_r \oplus \vec \xi)$.
The result will be some vector \mbox{$t \oplus \vec 0 \oplus \vec 0$}; by linearity, it follows that the operator 
\begin{equation}
	\tilde R \,:=\, \tau^{-2t} Z_r^{s} = \tau^{-2t} W_{s\unit_r}  \;,
\end{equation}
represented by the extended Pauli vector $t \oplus s\unit_r \oplus \vec \xi$, is the operator which is proportional to $Z_r^s$ and contained in $G_\sS$.
As in Section~\ref{sec:simTerminalMeas}, the coefficient $t$ must be a multiple of $s$ in order for $\tilde R$ to have a non-trivial $+1$-eigenspace.
Because $\tilde R$ commutes with $G_\sS$ and with $Z_r$\,, it stabilizes both the pre-measurement and the post-measurement state.
The operator $R = \tau^{-2h_{\ell+1}} Z_r$ which stabilizes $r$ after measurement satisfies $R^s = \tau^{-2hs} Z_r^{s} = \tilde R$; that is, the measurement outcome $h$ must satisfy $h_{\ell+1} \equiv t/\!s \pmod{\eta}$, or $h_{\ell+1} \in t/s + \eta\Z_d$.
We may show (see Appendix~\ref{apx:distributionMeasOutcomes}) that the outcome is in fact uniformly distributed among the residues modulo $d$ satisfying this constraint.

In the case $s = 1$ (\ie~if $\eta \in \{0,d\}$), the state is stabilized by some operator $\tau^{-2h} Z_r$ prior to measurement; the outcome is then a delta-peaked distribution, or ``uniformly distributed'' across the coset ${h + \eta\Z_d} = \{h\} \subset \Z_d$.
Furthermore, the measurement of $Z_r$ does not affect the state of the system; as an alternative to the above, we may transform the tableau to explicitly represent the fact that some operator $\tau^{-2h} Z_r$ stabilizes the pre-measurement state by column-reduction, using the operations for column combinations and reduction modulo $d$ of Sections~\ref{sec:defExtendedTableaus} and~\ref{sec:tableauReductionMod-d}.
The resulting tableau will contain a column of the form $h \oplus \unit_r \oplus \tilde{\vec \xi}$ explicitly representing the operator $\tau^{-2h} Z_r$.

\subsubsection{Computing and transforming between post-measurement states}
\label{sec:computingTransformingPostMeas}

Having fixed a given outcome $h_{\ell+1}$ of the $Z_r$ measurement, we perform the following transformations on $\bar T_{\tilde S}$ to represent the post-measurement state.
As the state is unchanged if $Z_r$ commutes with all stabilizer generators, we restrict ourselves to the case $s > 1$.

As $\tilde S_\ell$ does not commute with $Z_r$, it does not stabilize the post-measurement state.
But by construction, $\tilde S_\ell^{s}$ does.
To represent this, we multiply the $\ell\textsuperscript{th}$ column of $\bar T_{\tilde S}$ by $s$.
To keep the phase correction block consistent, we multiply the $\ell\textsuperscript{th}$ row of the phase correction block by $s$ as well.
We then augment the generating set to include the operator $R$ stabilizing the measured qudit: we do so by replacing the adjoined $(\ell+1)\textsuperscript{st}$ column of the tableau with $h_{\ell+1} \oplus \unit_r \oplus \vec \xi$\,.
(Note that by construction, $\vec \xi$ as given in Eq.~\eqref{eqn:measurementObservablePhaseCorrectionVector} is also the appropriate phase correction vector in this case).
We must also extend the phase-correction block by an additional row to represent the new generator of the stabilizer group; by symmetry, this is simply $\frac{d}{2} \unit_\ell \trans$ if $d$ is even, or $\vec 0\!\trans$ for $d$ odd.
This yields a new extended tableau $\bar T_{\sS'}$ representing a set $\sS'$ of stabilizer generators for the new state.

It may occur that the operator $\tilde S_\ell^{s}$ represented by the $\ell\textsuperscript{th}$ column of $\bar T_{\sS'}$ is in fact $\idop$, having a Weyl block consisting only of multiples of $d$.
This will occur for instance if $\eta = 1$ (which always occurs in the familiar case of $d$ prime).
If this is the case, we may drop the $\ell\textsuperscript{th}$ column of the new tableau entirely, and over-write the $\ell\textsuperscript{th}$ row of the phase correction block with the $(\ell+1)\textsuperscript{st}$ row rather than filling a new row.
More generally, it may be the case that $\tilde S_\ell^s$ can be generated by the other operators $\{S'_1, \ldots, S'_{\ell-1}, S'_{\ell+1}\} \subset \sS'$; this necessarily occurs when $\ell = 2n$, as this is the maximum number of operators needed to generate the group $G_{\sS'}$.
To discover whether this is the case, we may attempt to clear the Weyl block of the $\ell\textsuperscript{th}$ column of $T_{\sS'}$ modulo $d$ (\ie~reduce it to a $\ens{0,d}$-vector), as always using column operations with phase corrections as in Eq.~\eqref{eqn:extendedTableauRecombine}.
If successful, the resulting vector in the $\ell\textsuperscript{th}$ column represents the identity, and may be discarded as described above.

To simulate the transformation which occurs for a particular measurement outcome $h^\star \in t/s + \eta\Z_d$, we may simply set $h_{\ell+1} := h^\star$ by fiat, and perform the operations as above.
In a physical setting with actual stabilizer circuits acting on qudits of dimension $d$, as in the familiar case $d = 2$, one may simulate fixing the outcome by transforming the post-measurement state unitarily between the possible outcomes.
For $d \ge 2$ arbitrary, we may show that it suffices to act on the post-measurement state with the operator $\tilde S_\ell$, which is represented by conjugating each of the elements of $\sS'$ by $\tilde S_\ell$.
By construction, this operator  commutes with the generators $S'_j = \tilde S_j$ for $1 \le j < \ell$, as well as the generator $S'_\ell = \tilde S_\ell^{s}$, which are represented by the first $\ell$ columns of the tableau $\bar T_{\sS'}$; it only fails to commute with the generator $S'_{\ell+1} = R \propto Z_r$, transforming it to $R' = \tau^{-2\eta} R = \tau^{-2(h_{\ell+1} + \eta)} Z_r$ instead.
Repeated applications of powers of $\tilde S_\ell$ then suffice to produce any desired post-measurement state by adding a suitable multiple of $\eta$ to the exponent in the phase.

The entirety of the above analysis, for non-commuting $Z_r$ measurements, generalizes the well-known stabilizer formalism in prime dimension (as in Ref.~\cite{G98}).
If $d$ is prime, we have $\eta = 1$ as we have noted above, so that the post-measurement state is no longer stabilized by $\tilde S_\ell$ but instead by $R = \tau^{-2h} Z_r$, for some $h$ uniformly distributed over $\Z_d$. 
The novel features for $d$ arbitrary are that the possible outcomes over which $h$ varies can in general be any coset of the form $\kappa + \eta\Z_d$ (which is equal to the whole group $\Z_d$ only when $\eta = 1$), and that the post-measurement state is still stabilized by $\smash{\tilde S_\ell^{d/\eta}}$ (which is trivial only when $\eta = 1$).

\subsubsection{Special cases permitting simplified evolution under measurements}
\label{sec:extendedSimplif}

The extensions of the preceding sections to the stabilizer formalism allow us to simulate arbitrary stabilizer circuits for arbitrary dimensions $d \ge 2$.
We now briefly remark on simplifications which are possible in some special cases, which may lead to modest savings in the amount of effort and work-space required in practice.

When simulating a measurement of a Pauli operator $P = \tau^{-2\delta} W_{\vec p}$, it may be that no further operations depend on the measurement outcome (\ie~on the eigenvalue $\tau^{2h}$ corresponding to the post-measurement state); or if the outcome is used only to control Clifford operations, that representing the outcome by the state of a qudit provides no advantage.
In either case, the introduction of an explicit measurement register for controlled-Pauli operators to act upon is unnecessary.
We may compute the distribution of outcomes, and the post-measurement state conditioned on any particular outcome, in a similar way as described above but without the introduction of a measurement register.
Elaborating Eq.~\eqref{eqn:determineCommutingMeasurement}, we compute
\begin{align}
	\label{eqn:determineCommutingMeasurement'}
		\bm \phi
	\,&=\,
		(0 \oplus \vec p \oplus \vec 0)\!\trans \bar T_\sS   \;,
\end{align}
and consider whether it is equivalent to zero modulo $d$; if not, we consider what reversible column-transformation operations would map it to a row-vector of the form $\eta \unit_\ell\trans$, and then apply those same transformations to the tableau $T_\sS$.
Doing so yields a tableau representing a generating set $\tilde S = \{ \tilde S_1, \ldots, \tilde S_\ell \}$ in which only $\tilde S_\ell$ fails to commute with $P$; one may easily generalize the analysis above from that point on, substituting the Pauli vector $(0 \oplus \unit_r)$ representing $Z_r$ with the vector $(\delta \oplus \vec p)$ representing $P$.
Doing this is substantially similar to performing the same operations as in Section~\ref{sec:computingTransformingPostMeas}, as performed on a tableau where we \emph{have} simulated the $P$ measurement by a $Z_r$ measurement, with the primary difference being that we omit the additional column involved by explicitly introducing the ancilla $r$.

As noted in Section~\ref{sec:stabilizerTableaus}, extended stabilizer tableaus are unnecessary in the case of $d$ either prime or odd, as all stabilizer groups may be represented in those cases by proper stabilizer tableaus.
In the case of odd $d$, no special effort is necessary, as all stabilizer tableaus are proper in that case; we may omit phase correction blocks in that case.
In the remaining case $d = 2$ (\ie,~for qubits), further effort is required to ensure that the tableau of a post-measurement state is proper if we are to dispense with phase correction blocks.
If at least one of the elements of the stabilizer group $G_\sS$ fails to commute with the Pauli operator $P = \tau^{-2\delta} W_{\vec p}$ being measured, we may as usual obtain a generating set $\{\tilde S_1, \ldots, \tilde S_\ell\}$ in which exactly one generator $\tilde S_\ell$ does not commute with $P$; it suffices to compute a vector $\bar{\vec p} \equiv \vec p \pmod{2}$ such that $P = \tau^{-2\delta} W_{\vec p} = \tau^{-2\delta} W_{\bar{\vec p}}$, which is orthogonal modulo $D$ (where $D = 4$ in this case) to the columns of the Weyl block $\sfW_\sS$ representing $\tilde S_j$ for $1 \le j < \ell$.
We may compute $\bar{\vec p}$ using the techniques presented in Appendix~\ref{apx:makeProper} in this case, which should have a single solution.
If however $P$ commutes with the entire stabilizer group, there is no choice in how it may be represented in order to maintain a proper tableau: an operator proportional to $P$ is already generated by the group, and the system of equations in Appendix~\ref{apx:makeProper} determining a suitable representative for that operator has a unique solution $\bar{\vec p}$, for which $\sympl{\bar{\vec p}, \vec p} \equiv 2 \pmod{4}$ may hold.
In this case, however, no transformation of the state occurs upon measurement, so the existing proper tableau suffices to describe the post-measurement state.
We may then dispense with extended tableaus in these cases if desired (though extended tableaus still provide the benefit of making possible reduction of the coefficients modulo $2$ for tableaus over qubits).

Finally, for $d$ prime, certain elements of the analysis in Section~\ref{sec:simTerminalMeas} and Section~\ref{sec:evolutionUnderMeas} may be simplified to yield the known results for simulations of measurements in those cases~\cite{G98}.
For instance, in the case that not all generators $S_j \in \sS$ commute with the measurement operator $P$, it is not necessary to perform column transformations to obtain a column which represents an operator $\tilde S_\ell$ such that $[P,\tilde S_\ell] = \tau^{2\eta} = \tau^2$.
We may instead find any single operator $S_j$ which fails to commute with $P$, and compute some non-trivial power $S_j^t$ of it such that $[P,S_j^t] = \tau^2$, and use this to obtain a generating set in which only $S_j^t$ fails to commute with $P$ by combining other generators with appropriate powers of $S_j^t$.
Also, no non-trivial power of $S_j$ will be represented by the table $T_{\sS'}$ for the post-measurement state; in particular, the tableau will never have more than $n$ columns.

\section{Complexity of simulating stabilizer circuits}
\label{sec:complexitySimulate}

The main benefit provided by the formalism of this paper above the existing techniques in the literature is that computing phases are effectively reduced to simple linear transformations, thereby simplifying the individual steps of simulating stabilizer circuits, which should reduce the burden of carrying out transformations in analytical investigation and ad-hoc calculations.
(This is not to say that these operations are asymptotically more efficient: we remark on this distinction in Section~\ref{sec:simulateRunTimeComplexity}.)
However, the reduction to linear algebra in itself also makes certain complexity theoretic results easier to prove, when considering computational complexity classes which themselves are well characterized in linear algebraic terms.
Using the techniques of Sections~\ref{sec:simulateClifford} and~\ref{sec:fullStabilizerFormalism}, we generalize the results of Aaronson and Gottesman~\cite{AG04} concerning the complexity of simulating stabilizer circuits to qudits of arbitrary dimension. Specifically, natural decision problems concerning simulating stabilizer circuits on qudits of any fixed dimension $d$, involving at most a constant number of measurements, are complete for the complexity class \coMod[d]\Log\ consisting of problems which are log-space reducible to determining whether a system of equations mod $d$ is feasible~\cite{Beaudrap2012a}.
We describe these results in this section.

\subsection{Complexity of simulating unitary stabilizer circuits}
\label{sec:complexitySimulateUnitary}

We define \DefStabMeas\ to be the problem of deciding whether or not the state of a system of $n$ qudits of dimension $d$, initially in a computational state $\ket{\vec q}$ and then acted on by a unitary stabilizer circuit, is stabilized by some particular $P \in \cP_d\sox{n}$.
This is equivalent to the proposition that a measurement of the operator $P$ would yield a record of ``0'' with certainty, being a $+1$-eigenstate of $P$.
(Other possible outcomes $h \in \Z_d$ may be considered instead by testing whether $\tau^{-2h} P$ stabilizes the state.)
If the outcome ``0'' does not occur with certainty, it occurs either with probability $0$, or probability~$1/s$ (at most~$\frac{1}{2}$) for some integer $s$ which divides $d$.

In the special case of a $P = -Z_1$ measurement on qubits ($d = 2$), \DefStabMeas\ corresponds to the problem \GottKnill\ described by Aaronson and Gottesman~\cite{AG04}.
This problem belongs to the class $\parity\Log$ of problems which are log-space reducible to feasibility of systems of linear equations mod $2$ and verifying coefficients of matrix products mod $2$~\cite{D90}.
Relying on the result $\Log^{\parity\Log} = \parity\Log$~\cite{HSV00}, Aaronson and Gottesman describe an algorithm solving \GottKnill\ on a logspace machine with access to an $\parity\Log$ oracle, which is used to repeatedly simulate initial segments of the stabilizer circuit in order to compute the effects of the phase corrections induced on intermediate states of the circuit.
We show a more direct and generalized version of the result of Ref.~\cite{AG04} by characterizing the complexity of \DefStabMeas.

Just as \parity\Log\ is the class of problems which are log-space reducible to verifying coefficients of matrix products modulo $2$, we may define the class \coMod[d]\Log\ as the class of decision problems which are log-space reducible to verifying coefficients of matrix products modulo $d$~\cite{BDHM92} (see note~\footnote{%
	The class \coMod[d]\Log\ is usually defined as the class of decision problems for which there is a nondeterministic logspace Turing machine which accepts on a number of computational branches which is divisible by $d$ if and only if the input is a \emph{yes} instance.
	However, this precise definition is not particularly useful for our analysis.
	We will rely upon the characterization in terms of verifying coefficients of matrix products~\cite{BDHM92}, which is a standard approach in the literature to showing relationships of problems to \coMod[d]\Log.}).
In particular, this implies $\parity\Log = \coMod[2]\Log$.
As Ref.~\cite{Beaudrap2012a} shows, testing whether a system of equations is feasible mod $d$ is also a complete problem for \coMod[d]\Log\ for all $d \ge 2$; and in the particular case where $d$ is a prime power, \coMod[d]\Log\ may be characterized as those problems which are log-space reducible to \emph{evaluating} (as opposed to verifying) coefficients of matrix products modulo $d$.
These operations are in essence precisely what is required to simulate the transformations of a stabilizer tableau in a unitary stabilizer circuit, so that we may show:

\vspace*{-1ex}\theorem\
	\label{thm:defStabMeas}%
	For qudits of some fixed dimension $d \ge 2$, \DefStabMeas\ is complete for the class $\coMod[d]\Log$\,.
\proof\ 
As in the analysis of Section~\ref{sec:simTerminalMeas}, determining whether an $n$-qudit state, characterized by a (proper) stabilizer tableau $T_{\!f}$, is stabilized by an operator $P = \tau^{-2\delta} W_{\vec p}$ corresponds to determining whether there is a vector ${\delta' \oplus \vec p'}$ in the column-span of $T_{\!f}$ such that $P = \tau^{-2\delta'} W_{\vec p'}$.
This may be reduced to determining whether there is a solution to some system of equations $A \vec t \equiv (\delta \oplus p) \pmod{D}$ for some matrix $A$:
\begin{itemize}
\item
	For $d$ odd, in order for $P = \tau^{-2\delta'} W_{\vec p'}$, we require that $\vec p' - \vec p \equiv 0 \pmod{d}$, and that $2\delta' - 2\delta \equiv \sympl{\vec p',\vec p} \equiv 0 \pmod{d}$, so that $\delta \oplus \vec p$ is in the column span of $T_{\!f}$.
	We then set $A := T_{\!f}$.

\item
	For $d$ even, we still require $\vec p' - \vec p \equiv 0 \pmod{d}$, but this is no longer sufficient to ensure $\vec p = \vec p' \in \Z_D$; and again we require $2\delta' - 2\delta \equiv \sympl{\vec p',\vec p} \equiv 0 \pmod{D}$.
	Then $\vec p' - \vec p = d\vec x$ for some $\vec x \in \{0,1\}^{2n}$, and $\delta' = \delta + \tfrac{1}{2}\sympl{\vec p + d\vec x, \vec p} = \tfrac{d}{2}\sympl{\vec x, \vec p} \pmod{d}$.
	The vector $d \vec x$ corresponds to some representation of the identity operator $\idop = W_{d\vec x}$ which accounts for the difference in the phase coefficients, by the formula
	\begin{equation}
		\begin{aligned}[b]
				\tau^{-2\delta'} W_{\vec p'} \;&=\;	\tau^{-2\delta + \sympl{d\vec x,\vec p}} W_{\vec p + d\vec x} \\&=\; \tau^{-2\delta} W_{\vec p} W_{d\vec x}	\;,
		\end{aligned}
	\end{equation}
	by Lemma~\ref{lemma:WeylModular}.
	As in Section~\ref{sec:simTerminalMeas}, we introduce auxiliary columns of the form $\vec u_0 = d \oplus \vec 0$ and $\vec u_j := \frac{d}{2}\sympl{\unit_j, \vec p} \oplus d\unit_j$ for $1 \le j \le 2n$ to the tableau $T_{\!f}$, chosen so that $\vec u_j + (\delta \oplus \vec p)$ also represents the operator $P$ for each $0 \le j \le 2n$.
	Thus, it suffices to determine whether the system of equations $A \vec t = \mbox{$\bigl[ \vec u_0 \; \vec u_1 \; \cdots \; \vec u_{2n} \;\big|\; T_{\!f} \;\bigr]$} \vec t = (\delta \oplus \vec p)$ has solutions modulo $2d$.
\end{itemize}
In each case, to test the feasibility of such a system of equations has solutions with a \coMod[d]\Log\ algorithm, it is not necessary to store $T_{\!f}$ explicitly in the workspace; it suffices to be able to efficiently query individual coefficients of $T_{\!f}$ on demand, using only $O(\log(n))$ workspace.

In the case where $T_{\!f}$ represents a stabilizer state $U \ket{\vec q}$ obtained by acting on a standard basis state $\ket{\vec q} \in \cH_d\sox{n}$ with a Clifford operator $U \propto U_N \cdots U_2 U_1$, we define $T_{\!f}$ as the action of a sequence of operators $\bar C_N \cdots \bar C_2 \bar C_1$ acting on an initial tableau of the form $T_0 := \bigl[\, \vec q \;\big|\; 0 \;\bigr]\trans$, where each $\bar C_j$ is a conjugation tableau as in Eq.~\eqref{eqn:ConjTableau}.
We then consider two cases:
\pagebreak
\begin{itemize}
\item 
	Suppose $d = p^e$ for some prime $p$.
	Using the techniques of Ref.~\cite{Beaudrap2012a}, for any prime power $p^e$, we may compute the coefficients of such a matrix product modulo $p^e$ as part of a \coMod[p]\Log\ algorithm.
	Determining whether the system of equations $A \vec t \equiv \vec (\delta \oplus \vec p)$ is feasible modulo each prime-power divisor $p^e$ is thus in \coMod[p]\Log.

\item
	For qudit dimensions having a prime power decomposition	$d = p_1^{e_1} \cdots p_\ell^{e_\ell}$ for $\ell > 1$, note that $A \vec t \equiv (\delta \oplus \vec p) \pmod{D}$ is feasible if and only if $A \vec t \equiv (\delta \oplus \vec p) \pmod{p^e}$ is also feasible for every prime-power factor $p^e$ of $D$.
	Define the problem $\mathbf L_d$ to be the problem \DefStabMeas\ for a fixed qudit dimension $d$.
	Then we may characterize $\mathbf L_d$ as 
	\begin{equation}
		\mathbf L_d \;=\;	\mathbf L_{p_1^{e_1}} \;\cap\; \mathbf L_{p_2^{e_2}} \;\cap\; \cdots \;\cap\; \mathbf L_{p_\ell^{e_\ell}}	\;.
	\end{equation}
	By a standard normal form for \coMod[d]\Log~\cite[Prop.~3]{Beaudrap2012a}, it follows that $\mathbf L_d \in \coMod[d]\Log$ if and only if $\mathbf L_{p^e} \in \coMod[p^e]\Log\ ( = \coMod[p]\Log)$ for each prime power $p^e$ which divides $d$.
\end{itemize}
Thus, \DefStabMeas\ for qudits of dimension $d$, is in \coMod[d]\Log\ for all integers $d \ge 2$.
Finally, as simulating circuits consisting only of $\cX$ gates on qudits of dimension $d$ is itself \coMod[d]\Log-hard~\footnote{%
	Simulating networks made of reversible addition gates (that is, $\cX$ circuits) on tuples over $\Z_d$ is a hard problem for \coMod[d]\Log; this may be shown by a reduction from matrix powering, using the standard reduction from matrix powering to matrix inversion described by Cook~\cite{Cook85} and decomposing the upper-triangular matrices involved into elementary row operations, which is precisely how $\cX$ acts on standard basis states.},
it follows that \DefStabMeas\ for dimension-$d$ qudits is \coMod[d]\Log-complete.
\QED\endproof

We may generalize further, to consider the complexity of the problem of computing the output distribution of a measurement of a given observable.
For a given qudit dimension $d$ and a probability distribution over $\Z_d$\,, define \StabMeas\ to be the problem of deciding whether this distribution can be produced by measuring a given observable $P = \tau^{-2\delta} W_{\vec p}$ on the state produced by a specified unitary stabilizer circuit $U$ acting on a given standard basis state $\ket{\vec q}$.
We restrict the distributions taken as input to uniform distributions over a coset $\kappa + \eta \Z_d \subseteq \Z_d$, where $\kappa,\eta \in \Z_d$ are parameters specifying the distribution. (%
	The delta-peaked distributions of \DefStabMeas\ correspond to the case $\eta = d$, and testing whether $\tau^{-2\kappa} P$ stabilizes the state $U \ket{\vec q}$; the uniform distribution over all of $\Z_d$ corresponds to $\eta = 1$, with $\kappa$ being redundant.)
Consider the proper tableau $T_{\!\!\;f}$ just prior to measurement: by Section~\ref{sec:simTerminalMeas}, the outcome is uniformly distributed over $\kappa + \eta \Z_d \subseteq \Z_d$ if and only if the row-vector $\bm \phi := (0 \oplus \vec p)\trans T_{\!f}$ has the property that $\eta = \gcd(d,\phi_1, \phi_2, \ldots)$, and the integer vector $\frac{d}{\eta}[(\kappa - \delta) \oplus \vec p]$ is generated modulo $D$ by the columns of $T_{\!\!\;f}$ (together with auxiliary columns $\vec u_j := \tfrac{d}{2}\sympl{\unit_j,\vec p} \oplus d\unit_j$ and $\vec u_0 = d \oplus \vec 0$ in the case of $d$ even, as in the proof of Theorem~\ref{thm:defStabMeas}).
We may test both of these conditions by solving linear equations modulo $D$: this is obvious for the latter constraint, and we also have $\eta = \gcd(d,\phi_1, \phi_2, \ldots)$ if and only if the system of equations $\eta \equiv \bm\phi\trans \vec x \pmod{d}$ has solutions.
As \coMod[d]\Log\ is closed under logical conjunctions, \StabMeas\ is thus also complete for \coMod[d]\Log.

Note that \StabMeas\ is equivalent to determining whether the probability of obtaining a given outcome $h$ is equal to $1/s$ for some $s > 1$, as this holds if and only if the outcome is uniformly distributed over $h + (d/s)\Z_d$.  
We therefore have:
\theorem\
	For an initial state in the standard basis, any Pauli measurement observable $P \in \cP\sox{n}$, and any stabilizer circuit which performs a $P$ measurement, the following problems are all \coMod[d]\Log-complete: \bpar{a}~verifying predictions of a deterministic measurement outcome $h \in \Z_d$, \bpar{b}~verifying predictions of having probability $p$ of obtaining an outcome $h \in \Z_d$, and \bpar{c}~verifying predictions of the distribution of measurement outcomes.
\endtheorem

\noindent 
As a corollary, all of the problems described above may be simulated by $O(\log(n)^2)$-depth boolean circuits, as $\coMod[d]\Log \subseteq \NC[2]$~\cite{BDHM92}.

\subsection{The complexity of simulating stabilizer circuits with multiple measurements} 
\label{sec:complexitySimulateMeasurement}

We now consider the way in which the results of the preceding section extend to complexity containments for stabilizer circuits with measurements, \ie~in which the evolution of the state under measurement must be explicitly computed and where the outcomes may control further operations.
To simulate the transformation of a state under measurement --- as opposed to determining what the distribution of outcomes is, as in the \StabMeas\ problem --- we must describe how a \coMod[d]\Log\ algorithm might carry out the calculations described in Section~\ref{sec:evolutionUnderMeas}.

\subsubsection{Simulating evolution under a single measurement}

We first sketch an algorithm to verify any single coefficient of an (extended) post-measurement tableau $\bar T_{\sS'}$ in \coMod[p^e]\Log, for a qudit dimension of $p^e$ for some prime $p$, where the measurement outcome is somehow specified in the input and where we assume we may query coefficients of the pre-measurement tableau $\bar T_\sS$.
The tableau $\bar T_{\sS'}$ is that which results from a $Z_r$ measurement acting on a pre-measurement tableau $\bar T_\sS$ via the procedure of Section~\ref{sec:evolutionUnderMeas}.
The motivation for restricting to prime-power qudit dimension is to describe a solution involving query access to $\bar T_\sS$, which may be difficult to simulate in logarithmic space for composite $d$.
We indicate how this extends to composite qudit dimensions $d$, in a manner similar to the proof of Theorem~\ref{thm:defStabMeas}, before proceeding to the case of evolution under a sequence of measurements.

To evaluate a coefficient of the post-measurement tableau $\bar T_{\sS'}$, we do not have to store the entire tableau as it is transformed, so long as we can efficiently reconstruct the dependencies of the coefficient in question on the coefficients of the pre-measurement tableau $\bar T_\sS$ on demand.
We therefore describe how to reproduce which transformations are performed on $\bar T_\sS$, and the impact of these transformations on the desired coefficient.

The transformations performed on a tableau consist largely of column recombinations to clear the $(n{+}r{+}1)\textsuperscript{st}$ row of the tableau.
These may be performed by invertible transformations, in which one column is added or subtracted from another some number of times; for a prime-power modulus, determining a combination in which the $(n{+}r{+}1)\textsuperscript{st}$ coefficient of one of the columns is sent to zero is easy, using comparisons and divisions of fixed-size integers in $\{0,1,\ldots,D-1\}$.
We may consider consecutive pairs of columns in turn --- first considering combinations of the first column with the second column, then combinations of the second column with the third, and so on --- and determine for each pair the transformation which will clear the $(n{+}r{+}1)\textsuperscript{st}$ coefficient of the left-most column in each case.
By considering the effect of this sequence of column-combinations on the coefficients in other rows, we may determine how the coefficients in those rows transform, in order to determine what the value of any one given coefficient of $\bar T_{\sS'}$ would be.
We may do this using enough workspace to store the $(n{+}r{+}1)\textsuperscript{st}$ coefficients of whichever two columns we consider at each step of the algorithm, as well as the coefficients for the same two columns in any other row we require; and enough workspace to carry out simple calculations, such as division, on fixed-width integers.
Apart from the column recombinations of the tableau, computing $\eta$ (and the scalar factor $\alpha$ by which we multiply the final non-zero column, as described following Eq.~\eqref{eqn:columnReducedExtendedTableau}) can be easily performed in constant space, as can $s = d / \eta$ for $\eta > 0$.

In the case of a power of an odd prime, the above suffices to determine all the coefficients of the tableau $\bar T_{\tilde \sS}$ which represents the same pre-measurement state as $\bar T_\sS$, in which only a single generator $\tilde S_\ell$ fails to commute with $Z_r$.
In the case of $d$ even, the phase correction block and the phase vector both involve row operations for each column operation.
For a phase coefficient, we must also query off-diagonal coefficients from the phase correction block of the tableau, as it is being transformed; then some additional workspace is required to compute these coefficients.
A phase correction block coefficient itself may be subject to both column and row operations throughout the transformation of the stabilizer tableau: we may outline how these may be computed as follows.
Suppose we wish to compute the value that the $(h,j)$ coefficient of the phase correction block would have after performing the $k\textsuperscript{th}$ round of column recombinations, where $1 \le h,j,k \le 2n$.
The phase correction block is antisymmetric by construction, throughout the transformation of the tableau, which allows us to make the following observations:
\begin{enumerate}[label=\paritStyle{\roman*}]
\item
	If $j = h$, we return $0$; and if $j > h$, we may instead compute the negation of the $(j,h)$ coefficient of the phase correction block.
\item
	If $j < h$ and $k < h-1$, then none of the row-transformations on the phase correction block corresponding to recombining the first $k$ columns of the tableau have affected any coefficients in the $h\textsuperscript{th}$ row of the phase correction block, in which case we may simply simulate the effect of the first $k$ column combinations on the $(h,j)$ coefficient of the phase correction block, as for the Weyl block coefficients.
	(In particular, if $k < j-1$, we may simply return the corresponding coefficient of $\bar T_\sS$.)
\item
	If $j = k = h-1$, then the $(h,j)$ coefficient is in principle affected by row-transformations between the $(h-1)\textsuperscript{st}$ row and the $h\textsuperscript{th}$ row.
	However, as the $(h-1,h-1)$ coefficient of the phase correction block is zero, this row-operation has no effect, and we may reduce to the preceding case.
\item
	If $j < h \le k$, the $(h,j)$ coefficient of the phase block is affected by row-transformations which in general will have a non-trivial effect.
	We recursively compute coefficients $b_{t,j}$ corresponding to the $(t,j)$-coefficient of the phase correction block after $t$ column combinations, for $j+1 \le t \le h+1$.
	Starting by computing $b_{j+1,j}$ as in the preceding case, we compute each subsequent $b_{t,j}$ as follows: let $b'_{t,j}$ be the value of the $(t',j)$-coefficient of the phase correction block after the first $t-2$ column combinations, and then simulate the appropriate row-transformations with $b_{t-1,j}$ and $b'_{t,j}$ to compute the value of $b_{t,j}$ after the $(t-1)\textsuperscript{st}$ column combination.
\end{enumerate}
All of the above can be performed in constant workspace, using at most two levels of recursive evaluation of the coefficients of the phase-correction block (as in the final case above).

Having obtained a tableau $\bar T_{\tilde \sS}$ representing the pre-measurement group, for which only a single generator $\tilde S_\ell$ fails to commute with $Z_r$, we may easily describe the remaining calculations required to determine coefficients of the tableau.
The eigenvalue $\tau^{2t}$ of the pre-measurement state with respect to the operator $Z_r^s$ can be obtained by multiplying the measurement outcome specified for $Z_r$ at the input by $s$; we may then use a \coMod[p]\Log\ oracle to solve \DefStabMeas\ to determine whether the specified outcome is possible, and if so, proceed with the computation (perhaps returning an error value otherwise).
The column \mbox{$t \oplus \unit_r \oplus \vec\xi$} describing the stabilizer arising from the measurement will be the $(\ell+1)\textsuperscript{st}$ column of the tableau, for $\vec \xi \in \{0,\tfrac{d}{2},d,\tfrac{3d}{2}\}$ as described in Eq.~\eqref{eqn:measurementObservablePhaseCorrectionVector}, unless the column vector representing the generator $\tilde S_\ell^s$ can be expressed as a combination of \mbox{$t \oplus \unit_r \oplus \vec\xi$} together with the other columns, modulo $p^e$.
We may determine this once more with a \coMod[p]\Log\ oracle to solve systems of equations; and if there is indeed a solution, we omit the column corresponding to the old generator $\tilde S_\ell$ entirely.

As a minor variation of the procedure of Section~\ref{sec:evolutionUnderMeas}, we may switch the two columns representing the measurement stabilizer $\tau^{-2h} Z_r$ and the generator $\tilde S_\ell^s$, so that the former is always the $\ell\textsuperscript{th}$ column and the latter the $(\ell + 1)\textsuperscript{st}$ column in the case that it is non-trivial.
We may fixing the number of columns to the maximum of $2n$ as described in Section~\ref{sec:extendedTableauSize}, setting all additional columns to zero.
All of the above suffice to compute any particular coefficient of the post-measurement tableau $\bar T_{\sS'}$, modulo $p^e$; if a test-value for the coefficient is provided as input, we may then test congruence modulo $p^e$ as well.

Using the characterization of \coMod[d]\Log\ for arbitrary $d \ge 2$ (possibly divisible by multiple primes) described in Ref.~\cite[Prop.~3]{Beaudrap2012a}, this suffices to verify coefficients of a tableau transformed under measurement for arbitrary qudit dimension as well.
Any linear transformations which are invertible modulo the prime-power divisors $p_1^{e_1}$, $p_2^{e_2}$, \etc\ of $D$ are also invertible modulo $D$, by the Remainder Theorem; therefore each of the column transformations modulo the prime powers $p_j^{e_j}$ correspond to valid column transformations modulo $D$ as well.
By moving the column for the one generator of the group whose presence (or rather, whose status as a non-trivial generator) is uncertain to the $(\ell + 1)\textsuperscript{st}$ column, the columns in the tableaus for each prime-power divisor $p_j^{e_j}$ correspond to the same generators as one another; any prime-power divisor $p_j^{e_j}$ for which some generator corresponds to the zero vector merely represents a Pauli stabilizer which is proportional to some Weyl operator $W_{\!\smash{p_j}^{\!\!e_j}\;\!\!\vec v}$.
Thus, verifying the value of any given coefficient of a post-measurement tableau $\bar T_{\sS'}$ for qudits of any dimension $d \ge 2$ with respect to a $Z_r$ measurement is a problem contained in \coMod[d]\Log.

\subsubsection{Simulating multiple measurements in arbitrary dimensions}

From the foregoing, it is straightforward to use known oracle-closure results for prime-power $d$~\cite{HSV00} to show that a stabilizer circuit involving any constant number of measurements may be simulated in \coMod[d]\Log, for qudits of arbitrary dimension $d \ge 2$.
We proceed again along the same lines as the preceding section, by bounding the complexity for qudits of prime-power dimension, and then lifting to arbitrary dimension $d$.

In the procedure above for simulating a measurement of a tableau in prime-power dimension $p^e$, we assumed the ability to query individual coefficients of the tableau $\bar T_{\sS}$ which represents the pre-measurement state.
Each coefficient of the post-measurement tableau may be evaluated in \coMod[p]\Log, provided we supplement the computation with an oracle for the coefficients of $\bar T_{\sS}$.
If $\bar T_{\sS}$ arises from the simulation of a stabilizer circuit --- which may also be simulated in \coMod[p]\Log\ --- it follows that properties of interest of the post-measurement tableau may be computed in $\coMod[p]\Log^{\coMod[p]\Log}$.
If $\bar T_{\sS}$ itself arises from a circuit which involves a single measurement, an oracle for evaluating coefficients of $\bar T_{\sS}$ can be implemented using a $\coMod[p]\Log^{\coMod[p]\Log}$ oracle, \ie~an oracle which itself has access to an oracle to evaluate the coefficients of the tableau immediately following the first measurement.

Consider the problem of simulating a stabilizer circuit with $k > 1$ measurements, in the sense of computing coefficients of the stabilizer tableau (possibly in order to determine whether a given outcome $h_k \in \Z_d$ occurs for the final measurement), given some sequence of intermediate measurement outcomes $h_1, \ldots, h_{k-1} \in \Z_d$.
Generalizing the description above, the simplest way to regard the computational complexity of this problem is to provide one layer of nested \coMod[p]\Log\ oracles for each measurement, yielding a hierarchy of oracles for simulating each successive measurement.
Simulating a stabilizer circuit with $k$ measurements is then contained by the class
\begin{equation*}
	\coMod[p]\Log^{\coMod[p]\Log^{\coMod[p]\Log^{\ldots}}}  
\end{equation*}
with a tower of $k-1$ oracles to simulate the first $k-1$ measurements.
For any fixed $k > 1$ which is constant in the input size, this class is simply equal again to \coMod[p]\Log, by the oracle closure results of Ref.~\cite{HSV00} for $p$ prime (but see note~\footnote{%
	Note that no currently known techniques are known to simulate a tower of \coMod[p]\Log\ oracles of unbounded depth, on a \coMod[p]\Log\ machine; thus the distinction between $k \in O(1)$, and any number of measurements growing with the circuit size,  is important.}).
Thus simulating the outcomes of any fixed number of measurements can be simulated in \coMod[p]\Log\ for qudit dimension $p^e$.

For arbitrary dimensions, as we remark in the proof of Theorem~\ref{thm:defStabMeas}, the problem of simulating a stabilizer circuit on $d$ dimensional qudits can be reduced to the corresponding problem for each prime-power factor $p^e$ of $d$.
For instance, to determine whether a sequence of $k \in O(1)$ measurements yields outcomes $h_1, \ldots, h_k \in \Z_d$ with certainty, we may simulate the same circuit with all coefficients evaluated mod $p^e$, and also reduce each of the coefficients $h_t$ modulo $p^e$.
If instead we are interested in whether a sequence of outcomes $h_t$ arise with some probability (\eg~so that $h_1$ occurs with probability $p_1$, $h_2$ occurs with probability $p_{2|1}$ given that $h_1$ occurred, \etc), we may express the probabilities in terms of a product of powers $p^{-\gamma_t}$ for each prime divisor $p$ of $d$, and simulate for each measurement $1 \le t \le k$ whether or not the measurement outcome modulo $p^e$ occurs with probability $p^{-\gamma_t}$ by testing whether the measurement observable $Z_{r_t}^{p^{e-\gamma_t}}$ stabilizes the state at measurement.
As these problems may be contained in \coMod[p]\Log\ for prime $p$, and characterize the \emph{yes} instances for the same problems modulo $d$, these suffice to show:
\theorem\
	For any constant $k \ge 1$, for any stabilizer state given as input (described either as a standard basis state or via an initial stabilizer tableau), for any sequence $h_1, \ldots, h_k \in \Z_d$ of measurement outcomes, and for any sequence of conditional probabilities $p_1, \ldots, p_k$ of them occurring, the problem of determining whether the outcomes of the first $k$ measurements are the outcomes $h_j$ for $1 \le j \le k$, each with probability $p_1 p_2 \cdots p_j$, is \coMod[d]\Log-complete.
\endtheorem

\noindent\textbf{N.B.}~The issues described above relating to oracles do not arise if we are content to use a polynomial amount of workspace in the simulation. 
Simply using the techniques of the preceding sections, we may easily simulate arbitrary stabilizer circuits with any number of measurements on standard basis states in polynomial time, by transforming stabilizer tableaus which are explicitly stored in the work space.

\subsection{Run time complexity}
\label{sec:simulateRunTimeComplexity}

We conclude our considerations of the complexity of simulating stabilizer circuits with some remarks on more precise measures of complexity.

The computational complexity bounds of Sections~\ref{sec:complexitySimulateUnitary} and~\ref{sec:complexitySimulateMeasurement} do not impose any bounds whatsoever on the run-time complexity, except that it is polynomial by virtue of $\coMod[d]\Log \subseteq \NC[2] \subseteq \P$~\cite{BDHM92}.
In practice, the degree of this polynomial will be quite large, even for the non-deterministic logspace Turing machines~\cite{BDHM92} which are the usual model of computation used to define the class \coMod[d]\Log.
Furthermore, using the techniques of Ref.~\cite{Beaudrap2012a}, the degree of the run-time required to solve systems of linear equations modulo $d$ will increase with the value of $d$ --- and with the number of measurements being simulated --- according to the size of the prime-power factors of $d$.
However, this is an expected trade-off in time for the savings in the work-space required by the algorithms outlined in Sections~\ref{sec:complexitySimulateUnitary} and~\ref{sec:complexitySimulateMeasurement}.
As $\coMod[d]\Log \subseteq \NC[2] \subseteq \mathsf{DSPACE}(\log(n)^2)$, the large running-time may be regarded as a consequence of simulating stabilizer circuits with a constant number of rounds of measurement, but of any size, with either Turing machines which require only $O(\log(n)^2)$ space or polynomial-size boolean circuits of depth only $O(\log(n)^2)$.
The techniques used to show containment in \coMod[d]\Log\ may be regarded as demonstrating upper bounds on the abstract computational power of stabilizer circuits (for any fixed number of measurement rounds) for any $d \ge 2$;
a more practical approach to simulating stabilizer circuits, on a computer architecture having a small number of processors but a substantial amount of memory, is simply to maintain an explicit record of stabilizer tableaus and measurement outcomes.

Apart from the results of Sections~\ref{sec:complexitySimulateUnitary} and~\ref{sec:complexitySimulateMeasurement}, we may compare run-time complexity of the techniques of Sections~\ref{sec:simulateClifford} and~\ref{sec:fullStabilizerFormalism} to those of Ref.~\cite[Sec.~III]{AG04}, for simulating stabilizer circuits on qubits with the more traditional ``binary'' representation on qubits (\ie~for the case $d = 2$ alone).
In this setting, we consider the complexity only of simulating a generator of the Clifford group (such as a Pauli operator or a gate corresponding to one of the operators in Eq.~\eqref{eqn:spExamples}), or of single-qubit $Z$ measurements.
It must be noted that despite the elimination of quadratic phase corrections in the simulation of Clifford operations, there is no improvement in the asymptotic complexity of simulating a single Clifford group operation compared to the results of Ref.~\cite{AG04}; it is $O(n)$ in each case for a single gate acting on a tableau representing an $n$-qudit state (whether represented by a proper tableau, or an extended tableau).
Furthermore, the algorithm in Ref.~\cite{AG04} for simulating measurements in the case $d = 2$ is more efficient than the algorithm presented in this article for simulating measurements, as the procedure presented in Section~\ref{sec:complexitySimulateMeasurement} is essentially an extension of techniques of Ref.~\cite{GottPhD} to the case $d \ge 2$, for which Ref.~\cite{AG04} represents an improvement.
Indeed, it seems likely that the techniques of Ref.~\cite[Sec.~III]{AG04} could be easily extended using the linear formalism of this article to achieve a complexity of $O(n^2)$ for simulating $Z$ measurements in any fixed \emph{prime} dimension.
However, as those techniques seem to rely on the fact that $\Z_2$ (or $\Z_d$ for prime $d$) is a field, and in particular that the generating set for any stabilizer group has size at most $n$, it is not immediately clear how such techniques would extend even to the case of a prime power dimension.
The improvement of the formalism of this article over that of Ref.~\cite{AG04} is not in terms of run-time complexity, but rather extending the space-bounded complexity theoretic results to arbitrary dimensions $d \ge 2$, and exploiting the reduction to linear algebra to do this more directly in the case of prime powers.

\section{Conclusion}

We have presented techniques to simulate unitary stabilizer circuits on qudits of any constant dimension $d \ge 2$ using linear transformations, and simulate terminal measurements by solving systems of linear equations, using Weyl operators to represent Pauli operators.
In particular: we demonstrate that the Clifford group can in each case be described effectively using only Pauli operators and a group representation of the symplectic group $\Sp_{2n}(\Z_D)$ over the integers modulo $D$, where $D \in \{d,2d\}$ is determined according to whether $d$ is odd or even.
We also present the first explicit treatment of the evolution of a state under measurement for composite dimension, and demonstrate how this may be achieved by linear transformations of an extended tableau.
This leads to a simple, easy to use formalism for simulating stabilizer circuits on qudits of arbitrary dimension.

The motivation for the formalism of this article is not the actual run-time or circuit complexity, but rather a formalism for arbitrary dimension $d \ge 2$ which is as uniform as possible, in which simulating individual operations is as simple as possible (in the sense that each operation involves fewer natural arithmetic operations).
In doing so, we reduce the problem of simulation substantially to standard techniques of linear algebra, which is illustrated by the directness of the proof of Theorem~\ref{thm:defStabMeas} that simulating the measurement of a unitary stabilizer circuit is complete for the class \coMod[d]\Log\ for any $d \ge 2$ (extending the computational complexity results of Aaronson and Gottesman~\cite{AG04}).

There remain open questions with respect to the efficient simulation of stabilizer circuits, which tools of the sort presented in this article may help address.
We have shown that stabilizer circuits on qudits of any fixed dimension, and any size, may be simulated in \NC[2] provided it has a fixed number of measurements, as this task is complete for \coMod[d]\Log.
Is it possible to extend the number of measurements, for instance, to $O(\log(n))$, or (if we restrict the controlled-Clifford operations to controlled-Paulis) to $O(n)$?
Can similar results be obtained if we keep the number of measurements fixed, and perhaps even fix the size of the circuit, but allow the qudit dimension to be a prime power $p^e$ (of a fixed prime $p$) provided as input?
Finally, what bounds can we obtain for simulating stabilizer circuits in a distributed classical computational model based on linear transformations, such as linear network coding~\cite{ACLY00}?

\medskip

\subsection*{Acknowledgements}
\noindent
A substantial part of this work was performed while I was working for the \emph{Institut f\"ur Physik und Astronomie} at the Universit\"at Potsdam, Potsdam, Germany, with support from the EU (QESSENCE, MINOS, COMPAS) and the EURYI scheme.
The remainder was performed with support from the EC project QCS.

I would like to thank Earl Campbell, Matthias Ohliger, David Gross, Marcus Appleby, and the anonymous referees for helpful discussions and remarks on earlier drafts.

\bibliographystyle{apsrev4-1}
\bibliography{simpler-stab}

\appendix
\section{Proper stabilizer tableaus and symplectic Clifford transformations}

We now show that any generating set for a stabilizer group over $\cH_d\sox{n}$ has a proper stabilizer tableau, for $d$ prime or odd.
We also show how the same techniques apply to prove the ``lifting'' claim of Theorem~\ref{thm:symplecticClifford}.

\subsection{Proper stabilizer tableaus for $d$ prime and $d$ odd}
\label{apx:makeProper}

For $d$ either prime or odd, we wish to show that any sequence $S_1, \ldots, S_\ell \in \cP_d\sox{n}$ of commuting Pauli operators  (each with order at most $d$) may be represented by a sequence of vectors $\bar{\vec v}_1, \ldots, \bar{\vec v}_\ell \in \Z_D^{2n}$ and phases $\varphi_1, \ldots \varphi_\ell \in \Z_D$ such that $S_j = \tau^{-2\varphi_j} W_{\bar{\vec v}_j}$, and for which $\sympl{\bar{\vec v}_h, \bar{\vec v}_j} \equiv 0 \pmod{D}$ for all $0 \le h,j \le \ell$.
This is in fact trivially true in the case of $d$ odd: from the hypothesis that the operators $S_j$ commute pair-wise, it follows that $\sympl{\bar{\vec v}_h, \bar{\vec v}_j} \equiv 0 \pmod{D}$ by Lemma~\ref{lemma:WeylCalculus}
for any Pauli vectors $\varphi_j \oplus \bar{\vec v}_j$ representing the operators $S_j$ as above.
Thus, all stabilizer tableaus are proper for $d$ odd.

It remains to prove the result for the binary case $d = 2$.
More generally, for any even $d$, we may construct a proper tableau for the operators $S_j$, provided $S_1^{m_1} S_2^{m_2} \cdots S_\ell^{m_\ell} = \idop$ only if each $m_j$ is even. 
This holds, for example, for $\{ Z_1, Z_2, \cdots, Z_n \}$ on $n$ qudits, or any family of operators $S_j = U Z_j U\herm$ obtained from them by conjugation by $U \in \Cliff_n(d)$.
We set $\bar{\vec v}_1 = \vec v_1$, and for each subsequent $j > 1$, we construct $\bar{\vec v}_j = \bar{\vec v}_j + d\vec x_j$ for $\vec x_j \in \Z_D^{2n}$ which satisfies
\begin{subequations}
\label{eqn:iterativeProperConstraints}
\begin{alignat}{4}
	[ \vec v_j, \bar{\vec v}_j ] \;&\equiv\; 0		\!\!&\pmod{2d}	&\quad\text{and}	\\
	[ \bar{\vec v}_h, \bar{\vec v}_j ] \;&\equiv\; 0	\!\!&\pmod{2d}	&\quad\text{for all $1 \le h < j$}. 
\end{alignat}
\end{subequations}
We may construct such a vector $\bar{\vec v}_j$ as follows.
\begin{subequations}
The first constraint above ensures that $W_{\bar{\vec v}_j} = W_{\vec v_j}$\,.
As $\sympl{\vec v_j, \vec v_j} = 0$, we may simplify it to obtain $\sympl{\vec v_j, d\vec x_j} \equiv 0 \pmod{2d}$, or equivalently
\begin{align}
	\label{eqn:z-constraint-a}
  \sympl{\vec v_j, \vec x_j} \equiv 0 \pmod{2}.
\end{align}
For the second constraint above, note that as $\sympl{\vec v_h, \vec v_j}$ is a multiple of $d$ for each $1 \le h < j$, we have $\sympl{\bar{\vec v}_h, \vec v_j} = db_{h,j}$ for some $b_{h,j} \in \Z$; we may then expand this constraint to obtain
\begin{align}
	\!\!\!\!
	0 \;&\equiv\; \sympl{\bar{\vec v}_h, \bar{\vec v}_j } \,\equiv\, \sympl{\bar{\vec v}_h, \vec v_j} + \sympl{\bar{\vec v}_h, d\vec x_j}
	\notag\\&\equiv\;
		db_{h,j} + d\!\!\:\sympl{\vec v_h, \vec x_j} + 2d(\tfrac{d}{2})\sympl{\vec x_h, \vec x_j} \!\!
		\pmod{2d},
\end{align}
which we may further simplify to
\begin{align}
	\label{eqn:z-constraint-b}
	\sympl{\vec v_h, \vec x_j} \;&\equiv\;
		b_{h,j} 	\pmod{2}.
\end{align}
\end{subequations}
Let $\vec b_j = b_{1,j}\,\unit_1 + \cdots + b_{j-1,j}\, \unit_j$ and $\sfW_j = \bigl[\, \vec v_1 \,\, \cdots \,\, \vec v_j \,\bigr]$.
The constraints imposed by Eqs.~\eqref{eqn:z-constraint-a} and~\eqref{eqn:z-constraint-b} are then equivalent to
\begin{align}
	\label{eqn:solveZ}
		\sfW_j\trans \!\sigma_{2n} \,\vec x_j	\equiv	\vec b_j	\pmod{2}	\;.
\end{align}
Note that equations of the form $S_{1}^{m_1} \cdots S_{j-1}^{m_{j-1}} S_j^{-1} = \lambda \idop$ can only have solutions for $\lambda = 1$ in stabilizer groups; and we have specifically ruled out the possibility that we can obtain such an expression for $\idop$ using an odd exponent for $S_j$ in our hypotheses.
By Lemma~\ref{lemma:WeylModular}, it follows that the vectors $\vec v_j$ are linearly independent modulo $2$, in which case Eq.~\eqref{eqn:solveZ} is a solvable system of linear equations over $\Z_2$\,, with potentially many satisfactory solutions.
Selecting any one of them to fix a value of $\bar{\vec v}_j$, we then construct the next vector $\bar{\vec v}_{j+1}$, and so forth until we have obtained a sequence of Weyl vectors $\bar{\vec v}_1, \ldots, \bar{\vec v}_\ell$ forming a proper stabilizer tableau for the operators $S_1, \ldots, S_\ell$.

As we note above, the above technique can be applied for arbitrary qudit dimension $d$ as well for stabilizer tableaus whose Weyl blocks have independent columns.
For $d$ composite, not all tableaus have this property.
In particular, there exist stabilizer groups on $n$ qudits which have more than $n$ independent commuting generators, whose corresponding Weyl blocks therefore cannot have independent columns modulo $2$ or any other prime.
(See the beginning of Section~\ref{sec:extendedTableaus} for an example, and Section~\ref{sec:extendedTableauSize} for a more general bound on the number of independent generators.)

\subsection{Symplectic transformations performed by Clifford operations}
\label{apx:makeSymplectic}

We may apply similar techniques to the above to prove the ``lifting'' claim made in Theorem~\ref{thm:symplecticClifford}.
Note that if $W_{\vec v_j} = U W_{\unit_j} U\herm$ for $1 \le j \le 2n$, the operators $W_{\vec v_j}$ will all have order $d$, and will be independent as well by virtue of the independence of the operators $W_{\unit_j}$.
In particular, the vectors $\vec v_j$ will be linearly independent modulo $d$.
We may then apply the same lifting technique as above, except that rather than the constraints of Eq.~\eqref{eqn:iterativeProperConstraints}, we impose the constraints
\begin{subequations}
\begin{alignat}{4}
	[ \vec v_j, \bar{\vec v}_j ] \;&\equiv\; 0		\!\!\!&\pmod{2d}	&\;\;\text{and}	\\
	[ \bar{\vec v}_h, \bar{\vec v}_j ] \;&\equiv\; \delta_{h,n\!\!\;+\!\!\;j} - \delta_{j,n\!\!\;+\!\!\;h} 	\!\!\!&\pmod{2d}	&\;\;\text{for $h < j$},
\end{alignat}
\end{subequations}
where $\delta_{a,b}$ is the Kronecker delta.
These same congruences will already hold modulo $d$ (rather than $2d$) by the preservation of commutation relations from the conjugation by $U$.
By constructing each $\bar{\vec v}_j := \vec v_j + d\vec x_j$ as before, the same analysis may be applied to obtain constraints on each $\vec x_j$ sufficient to obtain the necessary inner products modulo $2d$ as well.
The matrix $\bar C = [\,\bar{\vec v}_1\,\,\cdots\,\,\bar{\vec v}_{2n}\,]$ that we obtain as a result is symplectic modulo~$D$.

\section{Supporting Lemmata concerning Weyl operators, stabilizer groups, and Pauli measurements}
\label{apx:spectraWeyl}

We now present additional technical properties of Weyl operators and stabilizer groups which generalize the known results for $d$ prime, to characterize of the evolution of stabilizer states upon measurement.
In particular, we show that this evolution is always uniformly random over a coset of the form $\kappa + \eta \Z_d$ for some $\kappa,\eta \in \Z_d$.
Much of this appendix describes results and techniques which are both simple and standard for the case $d = 2$ (see \eg~Ref.~\cite{GottPhD}) and for $d$ prime (Ref.~\cite{G98}).
However, such techniques do not seem to have been explicitly described for $d$ composite (for which technical obstacles exist, due \eg~to non-trivial operators $Z^a$ and $X^b$ which commute despite acting on a common qudit).

We again adopt the convention described at the beginning of Section~\ref{sec:simTerminalMeas}, that a measurement of a Pauli operator $P$ (of order at most $d$) stands for the measurement of some Hermitian operator $H$ with spectral diameter less than $2\pi$, such that $P = \exp(iH)$, and that the outcome is some $h \in \Z_d$ such that the result corresponds to the detection of a $\tau^{2a}$-eigenvector of $P$.
We also suppose all measurements to be non-destructive, leaving a residual system which is in an eigenstate of the observable $H$, and thus of the unitary operator $P$.

\subsection{Spectral properties of Weyl operators}

The following simple results about Weyl operators are helpful to demonstrate how stabilizer groups transform under measurements in Sections~\ref{apx:commutingPauliMeasurement} and~\ref{apx:noncommutingPauliMeasurement}, and may be of interest in the study of stabilizer codes over $\Z_d$.

As $W_{\vec v}^d = \idop$ for all $\vec v \in \Z^{2n}$, the eigenvalues of every Weyl operator are integer powers of $\tau^2$.
For $d$ composite, however, not all Weyl operators exhibit all possible powers of $\tau^2$ as eigenvalues.
(This is true for $d$ prime as well, if one includes the Weyl operator $W_{\vec 0} = \idop$.)
The eigenvalues that are exhibited by a Weyl operator $W_{\vec v}$, and their multiplicities, are governed in a simple way by the numerical relationships of the coefficients of $\vec v$.

\definition
 	\label{def:fundamentalVector}
	For a vector $\vec{v} \in \Z_D^m$, the \emph{harmonic number $\eta(\vec v) = \gcd(v_1, \ldots, v_m, d)$ of $\vec{v}$} is the largest positive integer $\eta \le d$ such that $\vec v \in \eta \Z_D^m$.
	We will say that $\vec{v}$ is \emph{fundamental} if $\eta(\vec v) = 1$.
\enddefinition
\noindent
Accounting for the harmonic number 
is necessary to bridge the gap with 
the case of $d$ prime~\footnote{
	Every non-zero vector in an actual vector space is fundamental, as $\gcd(x,p) = 1$ for $1 \le x < p$ and $p$ prime.
	This may be construed as precisely why stabilizer formalisms are easier to formulate in that case.}
to obtain constructions for all $d \ge 2$.
Weyl operators $W_{\vec v}$ with $\vec v$ fundamental have the greatest ability to ``distinguish'' between different states; 
Weyl operators other than these have fewer eigenvalues, and 
each eigenvalue has higher multiplicity, so that their eigenspaces decompose as the sum of multiple eigenspaces of some other Weyl operator.

\lemma
	\label{lemma:WeylFundamentalToZ}
	For each Weyl operator $W_{\vec v} \in \cP_d\sox{n}$, there exists a symplectic Clifford operator $U \in \sCliff_n(d)$ such that $U W_{\vec v} U\herm = Z_j^{\eta(\vec v)}$ for any $1 \le j \le n$.
\proof\
	For $\vec v \in \Z^{2n}$ fundamental, using the decomposition \eg~of Ref.~\cite[Sec.~IV]{HDM05}, one may show that there is a symplectic operation $C$ such that $C \vec v = \unit_n$\,.
	This construction consists essentially of using row-reductions on a stabilizer tableau to compute the greatest common divisor of the coefficients of $\vec v$; these row-reductions may be performed by left-multiplication by symplectic transformations.
	The operations needed to perform the row-reduction may be obtained by solving for coefficients $a_j$ such that $a_1 v_1 + \cdots + a_{2n} v_{2n} = 1$.
	By the construction described in the proof of Lemma~\ref{lemma:generateSymplClifford}, there then exists a symplectic Clifford $U \in \sCliff_n(d)$ such that $U W_{\vec v} U\herm = Z_n$ acting only on the $n\textsuperscript{th}$ qudit; we can then map this to any operator $Z_j$ by swaps.
	If $\vec v \in \Z^{2n}$ is not fundamental, let $\vec u = \vec v / \eta(\vec v)$; this vector is fundamental, so that there exists an operator $U$ such that $U W_{\vec u} U\herm = Z_j$.
	Then $U W_{\vec v} U\herm = U W_{\vec u}^{\eta(\vec v)} U\herm = Z_j^{\eta(\vec v)}$.
\QED\endproof

Note that the algorithm of Ref.~\cite[Sec.~IV]{HDM05} yields the described unitary $U$ in polynomial time, presented as a Clifford circuit of size $O(n \log(d))$.

\lemma
	\label{lemma:WeylEigenvalues}
	For any $n>0$ and $\vec v \in \Z^{2n}$, the eigenvalues of $W_{\vec v}$ are all of the integer powers of $\e^{2 \pi i \eta(\vec v) / d}$, each of which occurs with multiplicity $\eta(\vec v) \,d^{n-1}$.
\proof\
	By Lemma~\ref{lemma:WeylFundamentalToZ}, $W_{\vec v}$ has the same spectrum as $\smash{Z_n^{\eta(\vec v)}}$\!.
	Taking $Z_n$ as an operator acting on $\cH_d\sox{n}$, its spectrum consists of all of the integer powers of $\tau^2$ with multiplicity $d^{n-1}$ (that is, equal multiplicity).
	As $\eta(\vec v)$ divides $d$, we may show that the spectrum of $Z_n^{\vec v}$ is all of the $(d/\eta(\vec v))\textsuperscript{th}$ roots of unity also with equal multiplicity, which is to say $\eta(\vec v) d^{n-1}$; the same then holds for $W_{\vec v}$.
\QED

\corollary
	\label{cor:maximalWeylOrder}
	For any $\vec v \in \Z_D^{2n}$, the Weyl operator $W_{\vec v}$  has order $d/\eta(\vec v)$.
	In particular, it has order $d$ if and only if $\vec v$ is fundamental.

\subsection{Stabilizer groups of one-dimensional subspaces and commuting measurement observables}
\label{apx:commutingPauliMeasurement}

In this section, we prove the usual connection of Pauli observables which commute with every element of the stabilizer group with that measurement yielding a deterministic outcome, generalized to the setting of $d \ge 2$ arbitrary.
To do so, we prove a characterization (which is well-known for $d$ prime) of those stabilizer groups which describe unique pure states: they are maximal stabilizer groups, in the sense that any Pauli subgroup which strictly contains such a group is either nonabelian, or contains operators without $+1$ eigenvalues.
(Proofs in the case of $d$ prime typically make use of the fact that $\Z_d^{2n}$ is a field, which we cannot do for $d$ composite.)

\lemma
	\label{lemma:maximalStabilizer}
	Let $\sS = \ens{S_1, \ldots, S_\ell}$ be a generating set for a stabilizer group $G_\sS$ on $n$ qudits.
	The following are equivalent:\\[2pt]
	\begin{tabular}{c@{~~}l}
	\paritStyle{i}	&	
		$G_\sS$ stabilizes a unique state;	\\
	\paritStyle{ii}	&
		$G_\sS$ is a \emph{maximal} stabilizer group;	\\
	\paritStyle{iii}	&
		\parbox[t]{0.85\columnwidth}{$G_\sS$ is a stabilizer group of \emph{maximum size} (and has cardinality $d^n$).}
	\end{tabular}
\endlemma
\noindent
We note property \paritStyle{iii} above to emphasize the distinction from \paritStyle{ii}, and for the sake of completeness.
We will be interested primarily in certifying when a stabilizer group is inextensible, rather than when it has some particular cardinality.
\proof\
	Let $\sS \subset \cP_d\sox{n}$ be an arbitrary stabilizer group.
	For each generator $S_j \in \sS$, let
	\begin{align}
			\Pi_{S_j}
		\;=\;
			\frac{1}{d} \sum_{p=0}^{d-1}	\, S_{\!\!\;j}^{\;p}	
		\;=\;
			\frac{1}{|S_j|} \sum_{p=0}^{\!\!\!	|S_j|-1 \!\!\!}	S_{\!\!\:j}^{\;p}	\;,
	\end{align}
	where $|S_j|$ is the multiplicative order of $S_j$.
	It is easy to show that this operator projects onto the $+1$-eigenspace of $S_j$.
	Then $\Pi_\sS = \Pi_{S_1} \Pi_{S_2} \cdots \Pi_{S_\ell}$ projects onto the joint $+1$-eigenspace of $\sS$.
	By expanding each of the projectors $\Pi_j$\,, we may show that
	\begin{align}
		\label{eqn:stabilizedSpaceProj}
			\Pi_\sS
		\;=\;
			\frac{1}{|G_\sS|} \sum_{S \in G_\sS} \!S	\,,
	\end{align}
	as the terms in the sum run over all distinct combinations of powers $S_{1}^{\,p_1} S_{2}^{\,p_2} \cdots S_{\ell}^{\,p_\ell}$, generating each element of $G_\sS$.
	(As the group $G_\sS$ is the direct product of the cyclic groups generated by the operators $S_j$, it follows that $|G_\sS|$ is equal to the product of the orders $|S_j|$.)

	It is easy to show that $\paritStyle{i} \iff \paritStyle{iii}$: if $\sS$ stabilizes a unique state, it follows that $\Tr(\Pi_\sS) = 1$, so that
	\begin{align}
			1 \,=\, \Tr(\Pi_\sS) \,=\, \frac{1}{|G_\sS|} \sum_{S \in G_\sS} \!\Tr(S) \,=\, \frac{\Tr(\idop)}{|G_\sS|}\;,
	\end{align}
	where the final equality holds because $\idop$ itself is the only element of $G_\sS$ that has non-zero trace.
	Thus, $|G_\sS| = \Tr(\idop) = d^n$; and this cardinality is at a maximum, as $\Tr(\Pi_\sS)$ must be an integer.
	The converse is similar.

	It is also easy to show $\paritStyle{i} \implies \paritStyle{ii}$.
	Suppose $\sS$ stabilizes a unique state $\ket{\psi}$, which is to say that $\Tr(\Pi_\sS) = 1$.
	Let $P \in \cP_d\sox{n}$ be a Pauli operator which commutes with all of $\sS$.
	It is easy to show that $\sS$ also stabilizes $P \ket{\psi}$: by the uniqueness of $\ket{\psi}$, we therefore have $P \ket{\psi} = \lambda \ket{\psi}$ for some phase $\lambda$.
	Then $\lambda^{-1} P$ stabilizes $\ket{\psi}$, so that
	\begin{align}
			\label{eqn:pauliSum}
			\Pi_\sS	
		\;=\;
			\lambda^{-1} P \Pi_\sS
		\;=\;
			\frac{\lambda^{-1}}{|G_\sS|}
			\sum_{S \in G_\sS} P S \;.
	\end{align}
	By hypothesis, we then have	
	\begin{align}
			\frac{\lambda^{-1}}{|G_\sS|} \sum_{S \in G_\sS} \Tr(P S)
		\;=\;
			\Tr(\Pi_\sS)
		\;=\;
			1 \,.
	\end{align}
	As $\Tr(P S) \ne 0$ only if $PS \propto \idop$, this implies that there exists an operator $\bar S \in G_\sS$ such that $\bar S \propto P^{-1}$.
	As any two such operators would be proportional to one another, and can only be distinct if one of them failed to stabilize $\ket{\psi}$, such an operator is unique.
	Similarly, as $\bar S \propto \lambda P^{-1}$ both stabilize $\ket{\psi}$, these are equal as well.
	Then either $P \in G_\sS$, in the case that $\lambda = 1$; or the group obtained by extending $G_\sS$ by $P$ contains $\lambda \idop$ for $\lambda \ne 1$, and is therefore not a stabilizer group.
	Thus the uniqueness of the state $\ket{\psi}$ entails that $G_\sS$ is maximal as a stabilizer group in $\cP_d\sox{n}$.
	It remains to show that uniqueness of the stabilized state  is a necessary condition for maximality as a stabilizer group.

	Suppose instead that $G_\sS$ does not stabilize a unique state, and let $\ket{\psi_0}, \ket{\psi_1} \in \img(\Pi_\sS)$ be independent states stabilized by $G_\sS$.
	Consider the operator
	\begin{align}
		\Gamma = \ket{\psi_1}\bra{\psi_0} - \ket{\psi_0}\bra{\psi_1}\;.
	\end{align}
	As the Weyl operators span the set of operators on $\cH_d\sox{n}$ (by the corollary to Lemma~\ref{lemma:weylOrthogonal}, on page~\pageref{lemma:weylOrthogonal}), there exists some operator $W_{\vec v}$ such that $\gamma := \Tr(W_{\vec v}\herm \Gamma) \ne 0$.
	By construction, $\Gamma$ has trace zero; and as $\Gamma S = \Gamma = S \Gamma$ for every element $S \in G_\sS$, we have $\Tr(S\herm \Gamma) = 0$ for all $S \in G_\sS$ as well.
	Thus $W_{\vec v}$ is not proportional to any element of $G_\sS$.
	However, for any $W_{\vec s} \propto S \in G_\sS$, we have
	\begin{equation}
		\begin{aligned}[b]
		\gamma
	\;=\;
		\Tr\Bigl(W_{\vec v}\herm \Gamma\Bigr)
	\;&=\;
		\Tr\Bigl(W_{\vec v} \bigr( W_{\vec s} \Gamma W_{\vec s}\herm \bigr)\Bigr)
	\\&=\;
		\Tr\Bigl(\tau^{2\sympl{\vec v,\vec s}} W_{\vec v} \Gamma\Bigr)
	\;=\;
		\tau^{2\sympl{\vec v, \vec s}} \, \gamma \,,		  
		\end{aligned}
	\end{equation}
	where the penultimate equality holds by Lemma~\ref{lemma:WeylCalculus}.
	Then $\sympl{\vec v, \vec s} \equiv 0 \pmod{d}$ for all such $\vec s$, so that $S$ and $W_{\vec v}$ commute.
	It follows that $W_{\vec v}$ commutes with all of $G_\sS$.

	While $W_{\vec v}$ is not proportional to any element of $G_\sS$, there is a minimal integer $1 \le s \le d$ such that $W_{\vec v}^{s}$ is proportional to some $P \in G_\sS$\,.
	Let $\eta = \gcd(v_1, \ldots v_{2n}, d)$: 
	by Lemma~\ref{lemma:WeylEigenvalues}, $W_{\vec v}^s = W_{s\vec v}$ has eigenvalues consisting of integer powers of $\tau^{2s\eta}$ with equal multiplicity.
	As $P \propto W_{\vec v}^s$ has a non-trivial $+1$-eigenspace, the same is true for $P$.
	It follows that $P = \tau^{2s\eta r} W_{\vec v}^{s}$ for some $0 \le r < d$.
	
	Define $S_\ast := \tau^{2\eta r} W_{\vec v}$, which by construction satisfies $\Tr(S_\ast\herm \Gamma) = \tau^{-2\eta r} \Tr(W_{\vec v}\herm \Gamma) \ne 0$, and each of whose integer powers are either \textbf{(a)}~an element of $G_\sS$ or \textbf{(b)}~not proportional to any element of $G_\sS$.
	In particular, as $W_{\vec v}$ has order $d/\eta$ by construction, the only powers of $S_\ast$ which are proportional to the identity are in fact equal to $\idop$.
	Consider then the group $\overline G$ obtained by extending $G_{\bar \sS}$ by $S_\ast$\,.
	This group is abelian, as $S_\ast \propto W_{\vec v}$ commutes with all of $G_\sS$\,.
	Because the intersection of $\gen{S_\ast}$ and $G_\sS$ is the subgroup $\gen{S_\ast^s}$ by construction, we can decompose $\overline G$ into cosets of the form $S_\ast^t G_\sS$ for $0 \le t < s$.
	Define the operator
	\begin{align}
			\Pi_{\bar \sS}
		\;&=\;
			\frac{1}{|\overline G|} \sum_{S \in \overline G} S 	\;:
	\end{align}
	it is not difficult to show that $\Pi_{\bar \sS}^{2_{\phantom.}} = \Pi_{\bar \sS}$\,, so that this is a projection.
	Furthermore, as $S_\ast^0 \idop$ is the unique element of $\overline G$ proportional to the identity, we obtain
	\begin{align}
			\Tr(\Pi_{\bar \sS})
		\;&=\;
			\frac{1}{|\overline G|} \sum_{S \in \overline G} \Tr ( S )
		\;=\;
			\frac{\Tr(\idop)}{|\overline G|} > 0\,, \mspace{-25mu}
	\end{align}
	which implies that there exists a non-zero element of $\ket{\Psi} \in \img(\Pi_\sS)$ which is a $+1$-eigenvector of each element of $\overline G$.
	Thus $\overline G$ is a Pauli stabilizer group which strictly contains $G_\sS$, establishing $\paritStyle{ii} \implies \paritStyle{i}$.
\QED\endproof
\noindent
This result has an important consequence for Pauli measurements:
\corollary\
	Let $\ket{\psi} \in \cH_d\sox{n}$ be the unique $+1$-eigenstate of a stabilizer group $G_\sS \subset \cP_d\sox{n}$, and $P \in \cP_d\sox{n}$ an operator of order at most $d$ which commutes with every operator in $G_\sS$.
	Then $\ket{\psi}$ is undisturbed by measurements of $P$, and has a definite outcome $h$ such that $\tau^{-2h} P \in G_\sS$\,.
\proof\
	By the preceding Lemma, there exists a scalar $\lambda$ such that $\lambda P \in G_\sS$\,.
	Then $\ket{\psi}$ is an eigenvector of $P$ and is undisturbed by measurement.
	As $P$ has order at most $d$, we have $P = \tau^{-2\varphi} W_{\vec p}$ for some integer vector $\vec p \in \Z^{2n}$\,: its eigenvalues consist of integer powers of of $\tau^2$.
	The result of a $P$ measurement on $\ket{\psi}$ must be the $h \in \Z_d$ for which $\tau^{-2h} P \in G_\sS$.
\QED

\subsection{Measurement of observables which do not commute with all stabilizer generators}
\label{apx:noncommutingPauliMeasurement}
\label{apx:distributionMeasOutcomes}

Consider a maximal stabilizer group $G_\sS \subset \cP_d\sox{n}$ generated by a set of operators $\sS = \ens{S_1, \ldots, S_\ell}$, stabilizing a state $\ket{\psi}$.
We consider the effect on $\ket{\psi}$ of the measurement of an operator $P \propto W_{\vec p}$ such that $P^d = \idop$, but which does not commute with every $S_j \in \sS$.

\subsubsection{Reduction to the case of one generator not commuting with the measurement observable}

As in the case of $d$ prime, we may reduce to the case where at most one stabilizer generator fails to commute with the measurement observable by considering different generating sets of $G_\sS$\,.
For arbitrary $d \ge 2$, we may do this as follows.

For each $1 \le j \le \ell$, consider coefficients $0 \le \phi_j < d$ (not all zero) such that $P S_j P\herm = \tau^{2\phi_k} S_j$.
Note that the phases which are induced on Weyl operators $W_{\vec v}$ by commutation with $P$ depend on $\sympl{\vec p, \vec v}$, which is linear in~$\vec v$: thus the phases induced by commutation with $P$ on an arbitrary $S' \in G_\sS$ by commutation with $P$ is given by $m_1 \phi_1 + \cdots + m_\ell \phi_\ell$\,, where the integers $m_j$ are exponents such that $S' = S_1^{m_1} \cdots S_\ell^{m_\ell}$.
Let $\eta = \gcd(\phi_1, \ldots, \phi_\ell, d)$\,: there exist integer vectors $\vec x \in \Z^{\ell+1}$ such that
\begin{align}
		\eta \;=\; -z_{\ell\!\!\;+\!\!\;1} d \,+\, \sum_{j=1}^\ell z_j \phi_j	\;.
\end{align}
From the above remarks, there then exists an element $T = S_1^{z_1} \cdots S_\ell^{z_\ell} \in G_\sS$ for which $PTP\herm = \tau^{2\eta} T$.
As $\eta$ divides each coefficient $\phi_j$, let $\tilde \varphi_j = \phi_j / \eta$; we may then generate $G_\sS$ by the operators $\tilde \sS = \{\tilde S_0, \tilde S_1, \ldots, \tilde S_\ell \}$, setting $\tilde S_0 := T$ and $\tilde S_j := S_j T^{-\tilde \varphi_j}$ for $1 \le j \le \ell$.

\subsubsection{The subgroup of stabilizers commuting with the measurement}

In the above construction, each generator $\tilde S_j$ commutes with $P$ for $1 \le j \le \ell$.
In the case that $d$ is prime, they also generate the subgroup of $G_\sS$ which commutes with the observable $P$.
However, this does not hold for arbitrary $d$.
For example, consider a $Z_b$ measurement performed on a state $\ket{\psi}_{a,b}$ of a system of two qudits $a$ and $b$, where $d = d_1 d_2$ (for $d_1, d_2 > 1$) and where $\ket{\psi}$ is stabilized by
\begin{align}
		\ket{\psi}_{\:\!\!a,b} = \cX_{a,b}^{\!\:d_1} \, F_a \ket{0}_a \ket{0}_b\,.  
\end{align}
This state is stabilized by $\sS = \{Z_a^{-d_1} Z_b, X_a X_b^{d_1}\}$, as one may show by applying the transformations of Eq.~\eqref{eqn:specialCliffordGentorsMap} to the operators $\{Z_b\,,\, Z_a\herm\}$ stabilizing the state $\ket{0}_a \ket{0}_b$\,.
Let $S_1 = Z_a^{-d_1} Z_b$ and $S_2 = X_a X_b^{d_1}$.
Of the elements of $\sS$, only $S_2$ fails to commute with $Z_b$\,; and we obtain $\phi_1 = 0$, $\phi_2 = d_1$.
We may set $z_1 = 0$ and $z_2 = 1$ to obtain $\eta = d_1$, and from this define 
\begin{align}
	\begin{split}
	\tilde S_0 \,=\, T \,= \;	S_1^0 &S_2^1  \;\;=\, X_a X_b^{d_1}	\,,	\\
	\tilde S_1 \,= \;	S_1 &T^0 	\;\;=\, Z_a^{-d_1} Z_b	\,,	\\
	\tilde S_2 \,=  \;S_2 &T^{-1} =\,	\idop	\,.
	\end{split}
\end{align}
Note that $\{ \tilde S_1, \tilde S_2 \}$ alone fails to generate the operator $X_a^{d_2} = S_2^{d_2} \in G_\sS$, which commutes with $P$.

In the general case, we may characterize the subgroup of $G_\sS$ which commutes with $P$ as follows.
Consider an arbitrary operator $S' = \tilde S_0^{m_0} \tilde S_1^{m_1} \cdots \tilde S_\ell^{m_\ell} \in G_\sS$ which commutes with $P$.
As $\tilde S_j^{m_j}$ commutes with $P$ for each $1 \le j \le \ell$, $\tilde S_0^{m_0}$ must also commute with $P$, so that
\begin{align}
		\tilde S_0^{m_0}
	\;=\;
		P \tilde S_0^{m_0} P\herm
	\;=\;
		\tau^{2\eta m_0}\,\tilde S_0^{m_0}.
\end{align}
Then $m_0$ is a multiple of $d/\eta$, so that the operators $\sS_C = \{\tilde S_0^{d/\;\!\!\eta}, \tilde S_1, \tilde S_2, \ldots, \tilde S_\ell\}$ generate the subgroup of $G_\sS$ commuting with $P$.
(In the familiar case of $d$ prime, we necessarily have $\eta = 1$, so that $\smash{\tilde S_0^{d/\eta}} = \idop$ may be excluded as a generator; we then recover the result that $\tilde S_1, \ldots, \tilde S_\ell$ generate the commuting subgroup for $d$ prime.)

\subsubsection{Characterization of the post-measurement state and distribution of outcomes}

Following the above analysis, suppose that we have a generating set $\sS = \{ S_0, S_1, \ldots, S_\ell \}$ in which only $S_0$ fails to commute with $P$; specifically, we may suppose $P S_0 P\herm = \tau^{2\eta} S_0$ for some $\eta$ a divisor of $d$.
From the preceding section, $\sS_C = \{ S_0^{d/\eta}, S_1, S_2, \ldots, S_\ell \}$ generates the subgroup of $G_\sS$ which commutes with $P$.
The operators of $\sS_C$ also stabilize the post-measurement state: if we let
\begin{align}
	\label{eqn:resultProjector}
		\Pi_h	\;&=\;	\frac{1}{d} \sum_{j = 1}^d (\tau^{-2h} P)^j
\end{align}
be the projector onto the $\tau^{2h}$-eigenspace of $P$, we have
\begin{align}
		S \Pi_h \ket{\psi} = \Pi_h S \ket{\psi} = \Pi_h \ket{\psi}
\end{align}
for any $S \in G_{\sS_C}$.
The post-measurement state is naturally also stabilized by $R := \tau^{-2h} P = \tau^{-2u} W_{\vec p}$, where $h$ is the actual measurement outcome. 

To consider the distribution of possible values of $h \in \Z_d$, consider a Weyl operator $W_{\vec p} \propto P$, and let $u \in \Z_d$ be such that $R = \tau^{-2u} W_{\vec p}$.
By hypothesis, we have $W_{\vec p} S_0 W_{\vec p}\herm = \tau^{2\eta} S_0$, with $W_{\vec p} S_j W_{\vec p}\herm = S_j$ for all $1 \le j \le \ell$.
Let $s = d/\eta$; just as $S_0^{s}$ is the smallest power of $S_0$ which commutes with $P \propto W_{\vec p}$, we may show that
\begin{align}
  S_0 W_{\vec p}^s  S_0\herm	\;=\;	\bigl(\tau^{-2d} W_{\vec p}\bigr)^s \;=\; W_{\vec p}^s \,;
\end{align}
it follows that $W_{\vec p}^s$ commutes with all elements of $\sS$, and so is proportional to an element of $G_\sS$ by Lemma~\ref{lemma:maximalStabilizer}.
Specifically, it is proportional to an element of $G_{\sS_C}$, as $W_{\vec p}^s$ commutes with $P$.
The post-measurement state is then stabilized both by $\tau^{-2u} W_{\vec p}$ and by $W_{\vec p}^s$.
Thus the operator $(\tau^{-2u} W_{\vec p})^s W_{\vec p}^{-s} = \tau^{-2su} \idop$ stabilizes the post-measurement state, which implies that $su \equiv 0 \pmod{d}$, \ie~that $u$ is a multiple of $d/s = \eta$.
The result of measuring $P$ then satisfies $h \in \kappa + \eta\Z_d$, where $P = \tau^{-2\kappa} W_{\vec p}$.

The distribution of measurement outcomes within the coset $\kappa + \eta\Z_d$\,, and the relationships between them, straightforwardly generalize the standard results for $d$ prime.
Let $\Pi_h$ again be the projection onto the $\tau^{2h}$-eigenspace of $P$ as given in Eq.~\eqref{eqn:resultProjector}: then
\begin{align}
		S_0 \Pi_h
	\;\;&=\;\;
		\frac{1}{d} \sum_{j=1}^d S_0 (\tau^{-2h} P)^j
	\notag\\&=\;
		\frac{1}{d} \sum_{j=1}^d (\tau^{-2h - 2\eta} P)^j S_0
	\;\;=\;\;
		\Pi_{h+\eta} S_0 \,.
\end{align}
It follows then that $S_0 \Pi_h \ket{\psi} = \Pi_{h+\eta} \ket{\psi}$.
By repeated application of $S_0$, we can iterate through all the elements of $\kappa + \eta\Z_d$\,, from which it follows that \textbf{(a)}~all measurement outcomes in $\kappa + \eta\Z_d$ are equally likely, as the vectors $\Pi_{\kappa + \eta z} \ket{\psi}$ have the same norm; and \textbf{(b)}~the post-measurement state for the outcome $h$ can be transformed to that for $h + \eta z$ by performing the operation $S_0^z$.
Thus the ``eliminated'' stabilizer generator $S_0$ of the pre-measurement state becomes a unitary byproduct operator (in the sense of Ref.~\cite{RBB03}) which relates the possible post-measurement states; and the measurement outcome $h$ is uniformly distributed over the coset $\kappa + \eta\Z_d$, where $\tau^{2\kappa} P$ is a Weyl operator.

Note that the above analysis can accommodate the case of a deterministic measurement outcome, if we do not assume at the outset that $P$ does not commute with all of $G_\sS$.
If $W_{\vec p}$ commutes with all elements of $\sS$, it follows that $\eta = d$, in which case the solution set to Eq.~\eqref{eqn:measOutcomeCongruence} is a singleton modulo $d$: the distribution of outcomes is then the ``uniform distribution'' on that singleton set, \ie~a~delta distribution on the sole solution $h$.

As we have remarked previously, the familiar case for $d$ prime corresponds to $\eta = 1$ provided that $P$ does not commute with all elements of $G_\sS$.
Then Eq.~\eqref{eqn:measOutcomeCongruence} trivializes, imposing no constraints on the value of $h$.
The outcome is then uniformly distributed over all $h \in \Z_d$.
The generalization for $d$ composite is that the measurement outcome may in principle be uniformly distributed over a coset of any additive subgroup of $\Z_d$, and not just on some singleton set or on $\Z_d$ itself.

\end{document}